\documentclass[preprint]{JHEP3}
\usepackage{amssymb}

\usepackage{amsmath}
\usepackage[dvips]{graphicx}
\usepackage{epsfig}

\JHEPspecialurl{http://jhep.sissa.it/JOURNAL/JHEP3.tar.gz} 
\received{\today} 
\accepted{\today}

\title{Linear square-mass trajectories of radially and orbitally excited
hadrons in holographic QCD}

\author{Hilmar Forkel \\
Departamento de F\'isica, ITA-CTA, 12.228-900 S\~ao Jos\'e dos Campos, S\~ao
Paulo, Brazil and Institut f\"ur Theoretische Physik, Universit\"at
Heidelberg, D-69120 Heidelberg, Germany}
\author{Michael Beyer \\
Institut f\"ur Physik, Universit\"at Rostock, D-18051 Rostock, Germany}
\author{Tobias Frederico \\
Departamento de F\'isica, ITA-CTA, 12.228-900 S\~ao Jos\'e dos Campos, S\~ao
Paulo, Brazil}

\abstract{We consider a new approach towards constructing approximate 
holographic duals of QCD from experimental hadron properties. This 
framework allows us to derive a gravity dual which reproduces the 
empirically found linear square-mass trajectories of universal slope 
for radially and orbitally excited hadrons. Conformal symmetry breaking 
in the bulk is exclusively due to infrared deformations of the anti-de 
Sitter metric and governed by one free mass scale proportional to 
$\Lambda_{\rm QCD}$. The resulting background geometry exhibits dual 
signatures of confinement and provides the first examples of holographically 
generated linear trajectories in the baryon sector. The predictions for 
the light hadron spectrum include new relations between trajectory slopes 
and ground state masses and are in good overall agreement with experiment.}
\input{tcilatex}

\begin{document}

\section{Introduction \label{intro}}

Since the inception of Quantum Chromodynamics (QCD), progress in
understanding its low energy realm was hampered by the scarcity of adequate
techniques for handling strongly coupled Yang-Mills theories analytically.
The discovery of the AdS/CFT correspondence \cite{mal97,gub98} has given
promise for this situation\ to improve in a qualitative way. Indeed, the
ensuing dualities explicitly relate gauge theories at strong coupling to
physically equivalent string theories in ten-dimensional spacetimes which
become tractable at least in the weak (string) coupling and curvature
limits. These dualities are manifestations of the holographic principle \cite%
{holprin} and have triggered an entirely new way of thinking about
nonperturbative QCD.

The currently best understood dualities deal with supersymmetric and
conformal gauge theories, however, and so far the existence of an exact QCD
dual and its explicit form have not been established \emph{ab initio}. The
pioneering applications to QCD and hadron physics\footnote{%
Applications to remoter ``cousins'' of QCD, and in particular to their
glueball spectrum, have a longer history. See for example Refs. \cite{csa99}.%
} therefore relied on a minimal infrared (IR) deformation of the anti-de
Sitter (AdS) metric to model confinement \cite{pol02}. By restricting the
fifth dimension to a compact interval, this ``hard IR wall'' geometry softly
breaks conformal symmetry in a way consistent with high-energy QCD
phenomenology (including counting rules for exclusive scattering amplitudes
etc.) \cite{pol02}. The extension of this minimal approach into a search
program for the holographic dual of QCD, guided by experimental information
from the gauge theory side, is often referred to as AdS/QCD.

At present, this program is being pursued along two complementary lines. The
first one is mostly bottom-up: it assumes a five-dimensional, local
effective field theory in IR-deformed AdS$_{5}$ spacetime (and potentially
in additional background fields of stringy origin) to describe the gravity
dual of QCD and attempts to constrain its form and parameters by
experimental information. For various works in this direction see \cite%
{bos03,det05,erl05,dar05,kat06,kar06,bu} and references therein. The main
virtue of this approach lies in gathering and organizing information on
holographic QCD by using the wealth of accurate experimental data available
on the gauge theory side. The second type of approach is more directly
guided by the underlying, ten-dimensional brane anatomy of the gravity dual
and attempts to maintain closer ties to it\footnote{%
At present there is not enough calculational control over string theory in
curved spacetimes to allow for a strict top-down approach. Nevertheless, on
the maximally symmetric AdS$_{5}\times S^{5}$ a complete solution of the
world-sheet string theory may ultimately become feasible even in the
high-curvature regime \cite{ber05}.}. Up to now, however, this typically
comes at the price of less direct relations to QCD. For recent work along
these lines see for example \cite{td,tdhscatt,bro04} and references therein.

Over the last years AdS/QCD has met with considerable success in describing
hadron properties, the heavy quark potential \cite{mal98,kin00,and06},
vacuum condensates \cite{hir06}, QCD scattering amplitudes at high energy %
\cite{pol02,tdhscatt} etc. Many of these results were obtained on the basis
of the minimal hard wall implementation of IR effects. Although very useful
in several respects, the hard wall is also an oversimplification and has
revealed shortcomings when more complex and quantitative QCD properties are
considered. The perhaps most important limitation discovered so far is that
it predicts quadratic instead of linear square-mass trajectories both as a
function of spin and radial excitation quantum numbers~\cite%
{det05,kat06,bro06}, in contrast to experimental data and semiclassical
string model arguments (for highly excited states) which relate linear
trajectories to linear quark confinement \cite{shi05}.

Different ways to overcome this problem in the meson sector have recently
been proposed in Refs. \cite{kar06,and06,kru05}. As a result, linear Regge
trajectories $M^{2}\propto J$ for spin-$J$ excitations or the analogous
radial excitation trajectories $M^{2}\propto N$ of mesons could be
reproduced by different holographic models. Although the baryon sector \cite%
{det05,bar,hon07} exhibits similarly pronounced empirical trajectories of
the above type \cite{kle02}, however, a dual description for them has not
yet been found. Our primary goals in the present paper will therefore be to
understand how linear baryon trajectories can arise in the AdS/QCD
framework, and to establish dual descriptions for additional spectral
signatures of linear confinement.

To this end, we will focus on another striking and systematic feature in the
light hadron spectrum, namely the combined linear square-mass trajectories 
\begin{equation}
M^{2}=M_{0}^{2}+W\left( N+L\right)  \label{nltraj}
\end{equation}%
of Regge type on which radial ($N$) and orbital angular momentum ($L$)
excitations join. The trajectory structure (\ref{nltraj}) is experimentally
well established for both the light meson and baryon resonances. Two fits to
the meson data yield mutually consistent mean slopes $W=1.25\pm 0.15$~GeV$%
^{2}$~\cite{ani00} and $1.14\pm 0.013$~GeV$^{2}$~\cite{bug04}, respectively.
The fit to the light baryon resonances (i.e. to those consisting of up, down
and strange quarks) results in the somewhat smaller but still compatible
slope $W=1.081\pm 0.035$~GeV$^{2}$~\cite{kle02}. Hence the value $W\sim 1.1$
GeV$^{2}$ is approximately universal\footnote{%
This may be an indication for the conformal symmetry breaking scale $\propto
\Lambda _{\mathrm{QCD}}$ to be approximately hadron independent in the light
flavor sector, as noted for orbital excitations in Ref.~\cite{det05}.} for
all trajectories \cite{iac91}. The ground state masses $M_{0}$, on the other
hand, are channel dependent.

Our work will rely on the dual representation of hadronic states with higher
intrinsic angular momenta by metric fluctuations \cite{det05,bro04,bro06}
which suggests itself for our purposes because it provides direct access to
orbital excitations. The main strategy will be to devise and apply a method
for deriving IR deformed gravity duals which incorporate the experimental
information contained in the hadron trajectories (\ref{nltraj}). After a
brief summary of pertinent facts and results from AdS/QCD and the hard-wall
metric in Sec. \ref{ads}, we set out heuristically in Sec. \ref{ourpot} by
determining a minimal modification of the AdS mode dynamics which generates
the linear trajectories (\ref{nltraj}) in both meson and baryon spectra.
Subsequently, we explicitly construct the IR\ deformations of AdS$_{5}$
which encode the same dynamics, by deriving and solving differential
equations for the conformal symmetry breaking part of the warp factor in
Sec. \ref{metderiv}. Several new features of the resulting gravity
background are discussed in Sec. \ref{metdisc}. In Sec. \ref{phen} we
determine the value of the conformal breaking scale and compare the results
of our holographic description to experimental data. In Sec. \ref{sum},
finally, we conclude with a summary of our findings and mention a few
avenues for future improvements and applications.

\section{AdS/CFT correspondence and hadron spectrum \label{ads}}

The gauge/string duality~\cite{mal97,gub98} maps type IIB string theories in
curved, ten-dimensional spacetimes into gauges theories which live on the
(flat) 3+1 dimensional boundaries. For an UV-conformal gauge theory like
QCD, the dual string spacetime is the product of a five-dimensional
non-compact part which asymptotically (i.e. close to the boundary)
approaches the anti--de Sitter space $\mathrm{AdS}_{5}\left( R\right) $ of
curvature radius $R$, and a five-dimensional compact Einstein space $X_{5}$
(where $X_{5}=S^{5}\left( R\right) $ for the maximally supersymmetric gauge
theory) with the same intrinsic size scale. The line element therefore takes
the form~\cite{pol02}%
\begin{equation}
ds^{2}=e^{2A\left( z\right) }\frac{R^{2}}{z^{2}}\left( \eta _{\mu \nu
}dx^{\mu }dx^{\nu }-dz^{2}\right) +R^{2}ds_{X_{5}}^{2}  \label{metric}
\end{equation}%
(in conformal Poincar\'{e} coordinates) where $\eta _{\mu \nu }$ is the
four-dimensional Minkowski metric. Since $A\neq 0$ breaks conformal
invariance explicitly, one has to require $A(z)\rightarrow 0$ as $%
z\rightarrow 0$ in order to reproduce the conformal behavior of
asymptotically free gauge theories at high energies. The string modes%
\footnote{%
The dependence on the four dimensions $x$ and on the fifth dimension $z$
factorizes at least in the asymptotic AdS region.} $\phi _{i}\left(
x,z\right) =e^{-iP_{i}x}f_{i}\left( z\right) $ dual to physical states of
the gauge theory are particular solutions of the wave equations\footnote{%
The Klein-Gordon, Dirac and Rarita-Schwinger equations on AdS$_{5}$ are
discussed e.g. in Refs.~\cite{5dDirac,kir06}.} in the geometry (\ref{metric}%
) and fluctuations around it, and potentially in additional background
fields of stringy origin \cite{gub98}.

Casting these wave equations into the equivalent form of Sturm-Liouville
type eigenvalue problems, one finds 
\begin{equation}
\left[ -\partial _{z}^{2}+V_{M}\left( z\right) \right] \varphi \left(
z\right) =M_{M}^{2}\varphi \left( z\right)  \label{eveqm}
\end{equation}%
for the normalizable string modes $\varphi \left( z\right) =g\left( z\right)
f_{M}\left( z\right) $ dual to spin-0 ($M=S$) and spin-1 ($M=V$) mesons as
well as from the iterated Dirac and Rarita-Schwinger equations 
\begin{equation}
\left[ -\partial _{z}^{2}+V_{B,\pm }\left( z\right) \right] \psi _{\pm
}\left( z\right) =M_{B}^{2}\psi _{\pm }\left( z\right)  \label{eveqb}
\end{equation}%
for the string modes $\psi _{\pm }\left( z\right) =h\left( z\right) f_{B,\pm
}\left( z\right) $ dual to spin-1/2 and 3/2 baryons (where $\pm $ denote the
two chiralities of the fermions with $i\gamma ^{5}\psi _{\pm }=\pm \psi
_{\pm }$) \cite{det05}. The potentials $V_{M,B}$ contain all relevant
information on the string mode masses and the background metric (\ref{metric}%
) which also determines the functions $g\left( z\right) ,$ $h\left( z\right) 
$ introduced above. The eigenvalues $M_{M,B}^{2}$ constitute the mass
spectrum of the four-dimensional gauge theory on the AdS boundary. The
boundary conditions for the eigensolutions $\varphi $ and $\psi _{\pm }$ are
supplied by specifying the corresponding gauge theory operator according to
the AdS/CFT correspondence \cite{gub98} (cf. Eq. (\ref{bcgen})) and\ by the
requirement of normalizability (and minimal string action in case of
ambiguities) of the eigenmodes. In some cases a further, less well
determined boundary condition is imposed in the infrared, at $z=z_{m}$, in
order to break conformal symmetry.

The AdS/CFT correspondence establishes the link between the string mode
solutions of Eqs. (\ref{eveqm}), (\ref{eveqb}) and physical states on the
gauge theory side by prescribing an UV (i.e. $z\rightarrow 0$) boundary
condition for the solutions $f_{i}\left( z\right) $ of the five-dimensional
field equations \cite{gub98}. More specifically, for the dual of states $%
\left| i\right\rangle $ with four-dimensional spin 0 one has to select the
solution which behaves as $f_{i}\left( z\right) \overset{z\rightarrow 0}{%
\longrightarrow }z^{\Delta _{i}}$ where $\Delta _{i}$ is the conformal
dimension of the lowest-dimensional gauge theory operator which creates the
state $\left| i\right\rangle $. The wave functions of states with
four-dimensional spin $\sigma _{i}$ acquire an extra boost factor $%
z^{-\sigma _{i}}$, so that the boundary condition generalizes to%
\begin{equation}
f_{i}\left( z\right) \overset{z\rightarrow 0}{\longrightarrow }z^{\tau _{i}}%
\text{, \ \ \ \ \ \ }\tau _{i}=\Delta _{i}-\sigma _{i}  \label{bcgen}
\end{equation}%
where the scaling dimension $\Delta _{i}$ of the gauge-invariant operator is
replaced by its twist $\tau _{i}$ \cite{pol02}. The lightest string modes
are then associated with the leading twist operators, and therefore with the
valence quark content of the low-spin (i.e. spin 0, 1/2, 1, and 3/2) hadron
states~\cite{det05,bro06}. The duals of their orbital excitations (which
have no counterparts in the supergravity spectra) and hence of higher-spin
hadrons are identified with fluctuations about the AdS background \cite%
{det05,bro04}.

The leading-twist interpolators contain the minimal number of quark fields $%
q $ necessary to determine the valence Fock states of the hadrons. Their
intrinsic orbital angular momentum $L$ is created by symmetrized (traceless)
products of covariant derivatives $D_{\ell }$. This results in the operators 
$\mathcal{O}_{M,\bar{\tau}=L+2}=\bar{q}\Gamma D_{\{\ell _{1}}\dots D_{\ell
_{m}\}}q$ with $\Gamma =1,\gamma ^{5},\gamma ^{\mu }$ for scalar,
pseudoscalar and vector mesons, and $\mathcal{O}_{B,\bar{\tau}%
=L+3}=qD_{\{\ell _{1}}\dots D_{\ell _{q}}qD_{\ell _{q+1}}\dots D_{\ell
_{m}\}}q$ corresponding to spin-1/2 (or 3/2) baryons, with $L=\sum_{i}\ell
_{i}$. The boundary condition (\ref{bcgen}) is imposed on the solutions by
setting the values of the five-dimensional masses $m_{5,H}$ (which determine
the small-$z$ behavior) according to the twist dimension $\bar{\tau}$ of the
hadron interpolator to which they are dual \cite{gub98,pol02,bro06}, i.e.%
\begin{equation}
m_{5,H}R\ \rightarrow \left\{ 
\begin{tabular}{ll}
$\sqrt{\bar{\tau}(\bar{\tau}-d)}$ & $\text{for spin-0 mesons,}$ \\ 
$\sqrt{\bar{\tau}(\bar{\tau}-d)+d-1}$ & $\text{for vector mesons,}$ \\ 
$\bar{\tau}-2$ & $\text{for baryons,}$%
\end{tabular}%
\right. 
\begin{tabular}{l}
$\text{ \ \ \ }$ \\ 
$\text{ \ \ \ }$%
\end{tabular}
\label{adsbc}
\end{equation}%
where $d$ is the dimension of the boundary spacetime. The twist dimension of
the interpolating operators thereby enters the field equations and the
potentials $V_{M,B}$.

As outlined above, the duality of string modes to hadrons is based on their
association with the lowest-dimensional, gauge-invariant QCD interpolators
of matching quantum numbers. Hence this identification is incomplete as long
as fundamental quark flavor and chiral symmetry are not properly accounted
for. In the gravity dual, fundamental flavor arises from open string sectors
with the strings ending on added D-brane stacks \cite{kar02}. This mechanism
has an effective bulk description in terms of chiral gauge symmetries \cite%
{erl05,dar05} whose implementation into our framework will be left to future
work.

Nevertheless, a few comments on these issues and their impact especially on
the baryon sector may be useful already at the present stage. The simplest
and currently most popular top-down approach to quark flavor in the meson
sector introduces a D7-brane stack into (potentially deformed)\ AdS, with
open strings stretching between the D7- and D3-branes and with the duals of
flavored mesons living on their intersection \cite{kar02,kru03,bab04}.
Baryons require an additional D5-brane (wrapped around a compact part of the
ten-dimensional space) which contains baryonic vertices \cite{wit98} whose
attached strings pick up flavor by ending on the D7-branes as well \cite%
{kru03}. 

This construction is difficult to handle explicitly, however, and the
simpler propagation of fermionic fluctuations on the D3/D7 intersection 
\emph{without} the D5-brane was therefore recently studied as a preparatory
step \cite{kir06}. The resulting modes are dual to the superpartners of the
mesons investigated in Ref. \cite{kru03}, i.e. they correspond to fermionic
bound states of fundamental scalars (squarks) and fundamental as well as
adjoint spinors. These modes are solutions of Dirac equations similar to the
one used in our framework but fall into degenerate ($\mathcal{N}=2$)
supermultiplets with their mesonic partners.

Although bottom-up duals of baryons (with the correct $O(N_{c})$ scaling
behavior of their masses) obey Dirac equations as well \cite{kir06,det04},
they differ from the above ``mesinos'' in their assignment to gauge-theory
operators. Consequently they are subject to other AdS/CFT boundary
conditions and generate different mass spectra, as expected on physical
grounds. Such bottom-up descriptions of baryon duals were considered in Ref. %
\cite{det04}, on which we partially rely here, and also integrated \cite%
{hon07} into the approach of Ref. \cite{erl05}.

Another interesting dual representation of baryons has recently emerged from
a D4/D8-brane construction for fundamental flavor and chiral symmetry \cite%
{bar}. This approach generates a Chern-Simons gauge theory in the bulk whose
instantons are dual to solitonic baryons of Skyrme type \cite{zah86}, i.e.
to collective excitations of the meson fields which carry topological baryon
number and have masses of $O(N_{c})$ as well. Since low-energy properties of
Skymions can be described by Dirac fields (as they appear e.g. in chiral
perturbation theory \cite{gas88}), this approach to baryon duals may in fact
be complimentary to those of Refs. \cite{hon07,kir06,det04} outlined above.

The conformal potentials induced by the pure AdS$_{5}$ metric\footnote{%
The modes dual to baryons originate from the ten-dimensional Dirac equation.
Hence the nonvanishing eigenvalue of the lowest-lying Kaluza-Klein (KK) mode
of the Dirac operator on the compact space $X_{5}$ \cite{cam96} adds to the
five-dimensional mode mass $m_{5,B}$ (cf. e.g. Ref. \cite{det05}). Since the
AdS/CFT boundary conditions replace the whole mass term by a function of the
twist dimension of the gauge theory operator to be sourced, however, the KK
eigenvalue will not appear explicitly in the final expressions and has
already been absorbed into $m_{5,B}$.} (i.e. by Eq. (\ref{metric}) with $%
A\equiv 0$) are proportional to $1/z^{2}$: 
\begin{eqnarray}
V_{M}^{\left( \text{AdS}\right) }\left( z\right)  &=&\left[ \frac{15}{4}%
-\left( d-1\right) \delta _{M,V}+m_{5,M}^{2}R^{2}\right] \frac{1}{z^{2}},
\label{vmads} \\
V_{B,\pm }^{\left( \text{AdS}\right) }\left( z\right)  &=&m_{5,B}R\left[
m_{5,B}R\mp 1\right] \frac{1}{z^{2}}.  \label{vbads}
\end{eqnarray}%
Hence the normalizable eigensolutions of Eqs. (\ref{eveqm}) and (\ref{eveqb}%
) are Bessel functions whose order and eigenvalues depend on the boundary
conditions for the solutions $f_{i}\left( z\right) $ of the field equations
(cf. Ref. \cite{det05}). The conformal invariance inherited from the AdS
metric, however, prevents these potentials from carrying direct information
on IR effects of QCD.

The simplest way to approximately implement such IR effects, and in
particular confinement, is to impose a Dirichlet boundary condition on the
string modes at a finite IR scale $z_{m}$. This approach is prevalent among
current bottom-up models and amounts to a sudden onset of conformal symmetry
breaking by a ``hard-wall'' horizon of the metric at the IR brane \cite%
{pol02}, i.e. 
\begin{equation}
e^{2A_{\text{hw}}\left( z\right) }=\theta \left( z_{m}-z\right) ,\,\ \ \ \ \
\ z_{m}=\Lambda _{\text{QCD}}^{-1},  \label{hw}
\end{equation}%
and reduces the five-dimensional, noncompact space to an AdS$_{5}$ slice.
Even this minimal implementation of non-conformal IR effects into
approximate QCD duals (with only one free parameter related to $\Lambda _{%
\text{QCD}}$) can already predict a remarkable amount of hadron physics, as
outlined in the introduction. The investigation of hadron spectra and wave
functions in the approach of Refs. \cite{det05,bro06}, in particular, gave a
good overall account of the angular momentum excitation spectra for both
mesons and baryons.

In view of its simplicity, however, it is not surprising that the hard wall
confinement also reveals shortcomings. In particular, it predicts the square
masses of radially and orbitally excited hadrons to grow quadratically with $%
N$ and $L\,$\ \cite{det05,kat06}, in conflict with the linear Regge-type
trajectories found experimentally\footnote{%
In the hard wall model the first radial excitations of light mesons and the
nucleon, identified by a node in the string mode, appear at masses of about
1.8 GeV and 1.85 GeV, respectively~\cite{det05}, and are therefore difficult
to reconcile with the experimental $\pi (1300)$, $\rho (1450)$ and Roper 
N(1440)$P_{11}$ resonances. These shortcomings are suspected to be artifacts
of the hard wall metric as well~\cite{det05}.} and expected from the
semiclassical treatment of simple, relativistic string models \cite{shi05}.
While more detailed implementations of conformal symmetry breaking were able
to resolve this problem in the meson sector \cite{kar06,and06,kru05}, linear
baryon trajectories have so far not been obtained from a gravity dual. As
mentioned in the introduction, this provides part of our motivation to
search for a holographic representation which reproduces linear trajectories
in the baryon sector as well. (Note, incidentally, that the approach of Ref. %
\cite{kar06}, at least in its simplest form where a dilaton $\Phi \left(
z\right) \propto z^{2}$ is solely responsible for conformal symmetry
breaking, will not lead to linear trajectories in the baryon sector since
the dilaton interaction can be factored out of the Dirac equation and hence
does not affect the baryon spectrum.)

\section{Linear trajectories of radially and orbitally excited hadrons from
AdS type potentials \label{ourpot}}

We are now going to develop a gravity dual which manifests soft conformal
symmetry breaking directly in the potentials and is capable of generating
linear trajectories in both meson and baryon spectra. To this end, we find
in the present section suitable potentials heuristically and show that they
indeed reproduce the trajectories (\ref{nltraj}). In the subsequent Sec. \ref%
{metderiv} we then construct the IR deformations of the metric (\ref{metric}%
) which encode them holographically.

A natural guess for the $z$ dependence of potentials $V_{M,B}^{\left( \text{%
LT}\right) }$ which are able to generate the linear trajectorial (LT)
structure (\ref{nltraj}) is that it should be of oscillator type in the
infrared (i.e. quadratically rising with $z$ for $z\rightarrow \infty $).
The more challenging question is how to realize this behavior in a universal
way, i.e. on the basis of just one \textit{a priori} free mass scale $%
\lambda $ and such that the same slope $W$ and $N+L$ dependence emerges in
both meson and baryon channels. It turns out that this can be achieved at
the level of the twist\ dimensions which enter the five-dimensional mass
terms according to Eq. (\ref{adsbc}) after imposing the AdS/CFT boundary
conditions (\ref{bcgen}). Indeed, all the necessary information on conformal
symmetry breaking can be implemented into the AdS$_{5}$ potentials (\ref%
{vmads}), (\ref{vbads}) by replacing%
\begin{equation}
\bar{\tau}_{i}\rightarrow \bar{\tau}_{i}+\lambda ^{2}z^{2}.  \label{rrule}
\end{equation}%
(The hard wall restriction (\ref{hw}) of the AdS space becomes obsolete.)
The heuristic rule (\ref{rrule}) implies that the product of the
five-dimensional masses $m_{i}$ and the square root $a(z)=R/z$ of the AdS
warp factor grow linearly with $z$ for $z\rightarrow \infty $, thus
foreshadowing the linear trajectories (\ref{nltraj}) for both mesons and
baryons. The role of the hadron-independent mass scale $\lambda $ will
become more explicit below and in Sec. \ref{phen} where we relate it to the
trajectory slope $W$ and to the QCD scale.

As expected from \emph{soft} conformal symmetry breaking, the replacement (%
\ref{rrule}) does affect neither the $z\rightarrow 0$ behavior of the field
equations nor that of their solutions. Both the\ conformal symmetry on the
UV brane and the $z\rightarrow 0$ boundary conditions (\ref{bcgen}) from the
AdS/CFT dictionary are therefore preserved. As a consequence of the above
procedure, the mass terms in the pure AdS$_{5}$ potentials (\ref{vmads}), (%
\ref{vbads}) carry all information not only on the twist dimension (and thus
orbital excitation level) of the dual QCD operators but also on the
deviations from conformal behavior in the infrared. The underlying physical
picture will be discussed in Sec. \ref{metdisc}.

Recalling the expression $\bar{\tau}_{M}=L+2$ for the twist dimension of the
meson interpolators from Sec. \ref{ads} and making use of the replacement (%
\ref{rrule}) then turns the mesonic AdS potential (\ref{vmads}) into%
\begin{equation}
V_{M}^{\left( \text{LT}\right) }\left( z\right) =\left[ \left( \lambda
^{2}z^{2}+L\right) ^{2}-\frac{1}{4}\right] \frac{1}{z^{2}}  \label{vMconf}
\end{equation}%
(which holds for both spin 0 and 1) while the AdS potential (\ref{vbads}),
associated with the baryon\ interpolator of twist dimension $\bar{\tau}%
_{B}=L+3$, becomes 
\begin{equation}
V_{B,\pm }^{\left( \text{LT}\right) }\left( z\right) =\left\{ \left(
L+1\right) \left( L+1\mp 1\right) +\left[ 2\left( L+1\right) \pm 1\right]
\lambda ^{2}z^{2}+\lambda ^{4}z^{4}\right\} \frac{1}{z^{2}}.  \label{vBconf}
\end{equation}%
The normalizable solutions of the corresponding eigenvalue problems (\ref%
{eveqm}) and (\ref{eveqb}) can be found analytically. For the mesons one
obtains 
\begin{equation}
\varphi _{N,L}(z)=\mathcal{N}_{M;L,N}\text{ }\left( \lambda z\right)
^{L+1/2}e^{-\lambda ^{2}{z^{2}}/{2}}\text{ }{\text{L}}_{N}^{(L)}\left(
\lambda ^{2}z^{2}\right)  \label{phi}
\end{equation}%
where the ${\text{L}}_{N}^{(\alpha )}$ are generalized Laguerre polynomials %
\cite{abr72} and $\mathcal{N}_{H;L,N}$ are normalization constants. For the
(spin 1/2 and 3/2) baryons one similarly finds%
\begin{eqnarray}
\psi _{N,L,+}(z) &=&\mathcal{N}_{B+;L,N}\text{ }\left( \lambda z\right)
^{L+1}e^{-\lambda ^{2}{z^{2}}/{2}}{}\text{ }{\text{L}}_{N}^{(L+1/2)}\left(
\lambda ^{2}z^{2}\right) ,  \label{psi+} \\
\psi _{N,L,-}(z) &=&\mathcal{N}_{B-;L,N}\text{ }\left( \lambda z\right)
^{L+2}e^{-\lambda ^{2}{z^{2}}/{2}}{}\text{ }{\text{L}}_{N}^{(L+3/2)}\left(
\lambda ^{2}z^{2}\right) .  \label{psi-}
\end{eqnarray}%
Note that all eigenfunctions have appreciable support only over short
distances, in the small region $z\lesssim \sqrt{2}\lambda ^{-1}\simeq
\Lambda _{\text{QCD}}^{-1}$ (cf. Sec. \ref{phen}) close to the UV brane,
which is an expected consequence of confinement.

The corresponding eigenvalues 
\begin{eqnarray}
M_{M}^{2} &=&4\lambda ^{2}\left( N+L+\frac{1}{2}\right) ,  \label{mspec} \\
M_{B}^{2} &=&4\lambda ^{2}\left( N+L+\frac{3}{2}\right)  \label{bspec}
\end{eqnarray}%
show that the square masses of both mesons and baryons are indeed organized
into the observed $N+L$ trajectories. Moreover, the spectra (\ref{mspec})
and (\ref{bspec}) predict the universal slope 
\begin{equation}
W=4\lambda ^{2}  \label{slope}
\end{equation}%
for both meson and baryon trajectories in terms of the IR scale $\lambda $.
They also exhibit a mass gap (of order $\sqrt{W}$), another hallmark of
confining gauge theories, and the intercepts $M_{i,0}^{2}$ (cf. Eq. (\ref%
{nltraj})) relate the slope of the trajectories in a new way to their ground
state masses,%
\begin{eqnarray}
M_{M,0}^{2} &=&\frac{W}{2},  \label{mwmrel} \\
M_{B,0}^{2} &=&\frac{3W}{2}.  \label{mwbrel}
\end{eqnarray}%
The quantitative implications of these relations will be discussed in Sec. %
\ref{phen}.

\section{Derivation of the equivalent IR deformations of AdS$_{5}$ \label%
{metderiv}}

Although the existence of potentials (\ref{vMconf}) and (\ref{vBconf}) which
generate linear trajectories of the type (\ref{nltraj}) is encouraging, we
have not yet provided any dynamical justification for them. Indeed, for the
spectra (\ref{mspec}), (\ref{bspec}) to be the outcome of a dual gauge
theory, and for the hadronic quantum numbers to be associated with the
correct interpolating operators, one has to show that they emerge from
stringy fluctuations in a bulk gravity background. In the present section we
are going to establish this missing link by constructing the corresponding
background metric explicitly.

Of course, \textit{a priori} the existence of such a bulk geometry is far
from guaranteed, given the quasi \textit{ad-hoc} nature of the heuristic
rule (\ref{rrule}) which we used to find the potentials in the first place.
Moreover, it will prove sufficient to consider just the minimal set of
background fields\footnote{%
An effective $z$ dependence of the dual string mode masses may of course
also arise from additional background fields, as for example from a
Yukawa-coupled Higgs field~\cite{rin02}.}, consisting of the metric only. In
fact, we will show that even the simplest type of IR modifications of the AdS%
$_{5}$ metric, due to a non-conformal warp factor $e^{2A\left( z\right) }$
as anticipated in Eq. (\ref{metric}), can generate the potentials (\ref%
{vMconf}), (\ref{vBconf}). The success of this minimal approach can be at
least partially understood by noting that the potentials contain effects of
an order of magnitude which should arise from leading-order contributions to
the effective gravity action, i.e. from the metric. Higher-order
contributions due to dimensionful background fields (as e.g. the dilaton),
in contrast, would be suppressed by potentially large mass scales.

In order to prove the above assertions and to construct the non-conformal
warp factor, we first obtain the five-dimensional field equations for string
modes of the form $\phi \left( x,z\right) =f_{S}\left( z\right) e^{-iPx}$
(spin 0) and $V_{z}\left( x,z\right) =f_{V}\left( z\right) \varepsilon
_{z}e^{-iPx}$ (spin 1) which are dual to mesons and propagate in the
background of the metric (\ref{metric}) with an \textit{a priori}
unspecified warp function $A\left( z\right) $. The ensuing bulk equations
for the $z$ dependent part of the string modes are%
\begin{equation}
\left[ \partial _{z}^{2}+3\left( A^{\prime }-\frac{1}{z}\right) \partial
_{z}-\left( m_{5,S}R\frac{e^{A}}{z}\right) ^{2}+M^{2}\right] f_{S}\left(
z\right) =0
\end{equation}%
with $A^{\prime }\equiv \partial _{z}A$ and the four-dimensional invariant
square mass $M^{2}=P^{2}$, as well as 
\begin{equation}
\left[ \partial _{z}^{2}+3\left( A^{\prime }-\frac{1}{z}\right) \partial
_{z}+3\left( A^{\prime \prime }+\frac{1}{z^{2}}\right) -\left( m_{5,V}R\frac{%
e^{A}}{z}\right) ^{2}+M^{2}\right] f_{V}\left( z\right) =0.
\end{equation}%
The string modes dual to baryons can be
decomposed into left- and right-handed components,%
\begin{equation}
\Psi (x,z)=\left[ \frac{1+\gamma ^{5}}{2}f_{+}(z)+\frac{1-\gamma ^{5}}{2}%
f_{-}(z)\right] \Psi _{\left( 4\right) }(x),
\end{equation}%
where $\Psi _{\left( 4\right) }(x)$ satisfies the Dirac equation $(i\gamma
^{\mu }\partial _{\mu }-M)\Psi _{\left( 4\right) }(x)=0$ on the
four-dimensional (Minkowski) boundary spacetime. As a consequence, the
iterated five-dimensional Dirac equation for $\Psi $ reduces to 
\begin{eqnarray}
\left[ \partial _{z}^{2}+4\left( A^{\prime }-\frac{1}{z}\right) \partial
_{z}+2\left( A^{\prime \prime }+\frac{1}{z^{2}}\right) +4\left( A^{\prime }-%
\frac{1}{z}\right) ^{2}\right. \text{ \ \ \ } &&  \notag \\
-\left. \left( m_{5,B}R\frac{e^{A}}{z}\right) ^{2}\mp m_{5,B}R\frac{e^{A}}{z}%
\left( A^{\prime }-\frac{1}{z}\right) +M^{2}\right] f_{\pm }\left( z\right)
&=&0
\end{eqnarray}%
for the right and left handed modes with chiralities $i\gamma ^{5}f_{\pm
}=\pm f_{\pm }$.

These equations can be translated into the equivalent form of Schr\"{o}%
dinger-type eigenvalue problems (\ref{eveqm}) and (\ref{eveqb}) for $\varphi
\left( z\right) $ and $\psi _{\pm }\left( z\right) $ by writing $%
f_{S,V}\left( z\right) =\left( \lambda ze^{-A}\right) ^{3/2}\varphi \left(
z\right) $ and $f_{\pm }\left( z\right) =\left( \lambda ze^{-A}\right)
^{2}\psi _{\pm }\left( z\right) $ which eliminates the first-derivative
terms. The corresponding generalizations of the AdS$_{5}$ potentials (\ref%
{vmads}), (\ref{vbads}) are then read off from the eigenvalue equations as 
\begin{eqnarray}
V_{S}\left( z\right) &=&\frac{3}{2}\left[ A^{\prime \prime }+\frac{3}{2}%
A^{\prime 2}-3\frac{A^{\prime }}{z}+\frac{5}{2}\frac{1}{z^{2}}\right]
+m_{5,S}^{2}R^{2}\frac{e^{2A}}{z^{2}},  \label{vma} \\
V_{V}\left( z\right) &=&\frac{3}{2}\left[ -A^{\prime \prime }+\frac{3}{2}%
A^{\prime 2}-3\frac{A^{\prime }}{z}+\frac{1}{2}\frac{1}{z^{2}}\right]
+m_{5,V}^{2}R^{2}\frac{e^{2A}}{z^{2}}  \label{vva}
\end{eqnarray}%
and 
\begin{equation}
V_{B,\pm }\left( z\right) =m_{5,B}R\frac{e^{A}}{z}\left[ \pm \left(
A^{\prime }-\frac{1}{z}\right) +m_{5,B}R\frac{e^{A}}{z}\right] .  \label{vba}
\end{equation}%
The pure AdS$_{5}$ potentials are contained in these expressions for $%
A\equiv 0$. The AdS/CFT boundary condition, which relates the eigensolutions
to the dual meson (baryon) operators of twist dimension $\bar{\tau}_{M}=L+2$
($\bar{\tau}_{B}=L+3$), is imposed by adjusting the mass terms, 
\begin{eqnarray}
m_{5,S}^{2}R^{2} &=&\bar{\tau}_{M}(\bar{\tau}_{M}-4)=L^{2}-4,  \label{m2m} \\
m_{5,V}^{2}R^{2} &=&\bar{\tau}_{M}(\bar{\tau}_{M}-4)+3=L^{2}-1,  \label{m2v}
\\
m_{5,B}R &=&\bar{\tau}_{B}-2=L+1,  \label{mb}
\end{eqnarray}%
as outlined above. Equating the general potentials (\ref{vma}) - (\ref{vba})
to their heuristic counterparts (\ref{vMconf}), (\ref{vBconf}) leaves us
with differential equations for the corresponding warp functions. Their
solutions (subject to appropriate boundary conditions), finally, determine
the equivalent background metric of the form (\ref{metric}).

As already mentioned, it is \textit{a priori} uncertain whether there exists
an approximate gravity dual whose IR deformation can reproduce a given
five-dimensional potential and spectrum, simply because it may not result
from a boundary gauge theory. In the above approach this is reflected in the
fact that the nonlinear, inhomogeneous differential equations for $A\left(
z\right) $ may not have physically acceptable solutions. Our next task is
therefore to construct and analyze the solution spaces of these differential
equations for the heuristic potentials (\ref{vMconf}), (\ref{vBconf}).

\subsection{Baryon sector}

We begin our discussion in the baryon sector where the equation for $A$ is
of first order and hence has less and generally simpler solutions. Equating
the potential (\ref{vBconf}), whose representation in terms of the
background geometry we wish to construct, to the potential (\ref{vba}) for
general $A$ results in the nonlinear, inhomogeneous differential equation%
\begin{equation}
\pm \left( zA^{\prime }-1\right) +le^{A}-\left[ l\left( l\mp 1\right)
+\left( 2l\pm 1\right) \lambda ^{2}z^{2}+\lambda ^{4}z^{4}\right] \left(
le^{A}\right) ^{-1}=0  \label{bode}
\end{equation}%
of first order (where $l\equiv L+1$ and $\pm $ refers to the two baryon
chiralities) whose solution determines the non-conformal part $\exp \left[
2A_{B}\left( z\right) \right] $ of the equivalent warp factor.

Remarkably, the exact (and essentially unique) solution of Eq. (\ref{bode})
subject to the conformal boundary condition $A_{B}\left( 0\right) =0$ can be
found analytically and turns out to be%
\begin{equation}
A_{B}\left( z\right) =\ln \left( 1+\frac{\lambda ^{2}z^{2}}{L+1}\right) .
\label{bsoln}
\end{equation}%
Note that the same solution holds for both baryon chiralities. The leading
contribution to the non-conformal part of the warp factor at small $z^{2}\ll
\lambda ^{-2}$ is therefore%
\begin{equation}
e^{2A_{B}}=e^{\frac{2}{L+1}\lambda ^{2}z^{2}+O\left( \lambda
^{4}z^{4}\right) }
\end{equation}%
which has the form of the analogous warp factor $\exp \left( cz^{2}/2\right) 
$ used in Refs. \cite{and06,and206} (together with a constant dilaton)\ to
obtain a linear quark potential and a linear (mesonic) Regge trajectory.

\subsection{Scalar meson sector \label{sm}}

In the following we are going through the analogous construction for $A_{S}$
in the spin-0 meson sector, which will turn out to be more multi-faceted.
Equating the meson potential (\ref{vMconf}) to its general-$A$ counterpart (%
\ref{vma}) produces again a nonlinear, inhomogeneous equation, but for $%
A_{S} $ it is of second order: 
\begin{equation}
z^{2}A^{\prime \prime }+\frac{3}{2}\left( zA^{\prime }\right)
^{2}-3zA^{\prime }+\frac{2}{3}\left( L^{2}-4\right) \left( e^{2A}-1\right) -%
\frac{2}{3}\lambda ^{2}z^{2}\left( \lambda ^{2}z^{2}+2L\right) =0.
\label{mde}
\end{equation}%
In addition to $A_{S}\left( 0\right) =0$, its solutions require a second
boundary condition which as of yet remains unspecified and will be
determined below. This added freedom provides one of the reasons for the
solution space in the meson sector to be larger and more diverse than in the
baryon sector.

In addition, the $L$ dependence of the solutions $A_{S}$ is more
heterogeneous since the sign of the mass term $m_{5,S}^{2}R^{2}$ in the
field equation can be either negative, zero, or positive (cf. Eq. (\ref{m2m}%
)). These three\ cases generate qualitatively different solution behaviors.
The positive sign corresponds to $L>2$ and is associated with irrelevant
gauge theory operators according to the renormalization group (RG)
classification. The solutions for $L>2$ will turn out to be qualitatively
similar to those in the baryon sector. The massless case $m_{5,S}^{2}R^{2}=0$
corresponds to $L=2$ and to a marginal operator in the RG sense. The duals
of the lowest orbital excitations $L=0,1$, finally, represent relevant
operators and are tachyons\footnote{%
Scalar AdS$_{5}$ tachyons with masses satisfying the Breitenlohner-Freedman
bound $m^{2}R^{2}\geq -d^{2}/4$ (which includes the cases we encounter here
for $d=4$)\ do not cause instabilities, as has been known for some time \cite%
{bre82}.} with $m_{5,S}^{2}R^{2}=-4,-3$. In the following, we will discuss
these three cases in turn.

\paragraph{$L=0,1$:}

Due to their tachyonic nature, the $L=0,1$ solutions are perhaps the most
interesting ones. The negative mass term together with the specific
inhomogeneity generated by the potential (\ref{vMconf}) forces these
solutions to develop a singularity at finite $z=z_{m}$, which restricts the
spacetime to an AdS$_{5}$ slice\footnote{%
In our case this is a geometric consequence of requiring the holographic
dual to generate potentials which exhibit the linear spectral trajectories (%
\ref{mspec}), (\ref{bspec}). Alternatively, additional branes may restrict
the fifth dimension in the IR \cite{pol00}.}. The position and sign of these
singularities depends on the second boundary condition for $A_{S}\left(
z\right) $. At $z=0$ this boundary condition may e.g. be imposed\footnote{%
The analysis of the linearized approximation to Eq. (\ref{mde}) in App. \ref%
{linode} shows that $A_{S}\left( 0\right) =0$ automatically implies $%
A_{S}^{\prime }\left( 0\right) =0$, $A_{S}^{\prime \prime }\left( 0\right)
=2L\lambda ^{2}/\left( L^{2}-7\right) $ for $L\neq 0$ as well as $%
A_{S}^{\prime }\left( 0\right) =A_{S}^{\prime \prime }\left( 0\right)
=A_{S}^{\prime \prime \prime }\left( 0\right) =0$, $A_{S}^{\prime \prime
\prime \prime }\left( 0\right) =-3\lambda ^{4}$ for $L=0$, as a consequence
of the inhomogeneity.} on $A_{S}^{\prime \prime \prime }\left( 0\right) $
for $L\neq 0$ and on $A_{S}^{\prime \prime \prime \prime \prime }\left(
0\right) $ for $L=0$. The singularities have positive (negative) sign, i.e. $%
A\rightarrow \pm \infty $, if $A_{S}^{\prime \prime \prime }\left( 0\right) $
(or $A_{S}^{\prime \prime \prime \prime \prime }\left( 0\right) $ for $L=0$)
is chosen larger (smaller) than a critical value.

Independently of the sign of the singularities, furthermore, the warp factor%
\begin{equation}
\frac{R^{2}}{z^{2}}e^{2A_{S}\left( z\right) }\geq c\left( z_{m}\right) \geq 0
\end{equation}%
($c\left( z_{m}\right) =R^{2}\exp \left[ 2A_{S}\left( z_{m}\right) \right]
/z_{m}^{2}=0$ for negative singularities) of all solutions remains bounded
from below for all $z$ up to $z_{m}$ where the dual spacetime ends. The
above behavior provides a sufficient confinement criterion in the
five-dimensional holographic description.\ The Wilson loop \cite{mal98} then
shows an area law (since the strings are localized at $z_{m}$) and the gauge
theory develops the expected mass gap $M_{\min }\geq z_{m}^{-1}$ \cite{kin00}%
. In fact, the hard-wall horizon (\ref{hw}) may be considered as a simple
model for this type of behavior. It corresponds to the development of an
abrupt negative singularity of $A$ at $z_{m}$.

The origin and locus of the negative singularities can be understood
quantitatively by obtaining a series solution for $z\ll \sqrt{2}\lambda
^{-1} $ in the form%
\begin{equation}
A_{S}\left( z\right) =\frac{1}{2}\ln \left[ 1+\sum_{n=1}^{\infty
}A_{n}\left( L\right) \left( \frac{\lambda ^{2}z^{2}}{2}\right) ^{n}\right]
\label{amser}
\end{equation}%
which already incorporates both boundary conditions, i.e. $A\left( 0\right)
=0$ and a second one to become explicit below. The coefficients $A_{n}$ may
be calculated by inserting the expansion (\ref{amser}) into Eq. (\ref{mde}).
The first three are (for $L\neq 2$) 
\begin{equation}
A_{1}\left( L\right) =\frac{4L}{L^{2}-7},\text{ \ \ \ \ }A_{2}\left(
L\right) =\frac{4}{L^{2}-4}\left( 1-\frac{9}{4}A_{1}^{2}\right) ,\text{ \ \
\ \ }A_{3}\left( L\right) =\frac{-12}{L^{2}+5}A_{1}^{3}.
\end{equation}%
In general, the inhomogeneity of Eq. (\ref{mde}) determines the leading
small-$z$ behavior of the $A_{S}$, as revealed by the solutions (\ref{linL01}%
) - (\ref{lL3}) of the linearized equation given in App. \ref{linode}. For $%
L=0,1$, in particular, it forces the $z$ dependence inside the logarithm to
start out quadratically and yields 
\begin{eqnarray}
A_{S,L=0,1}\left( z\right) &=&\frac{1}{2}\ln \left[ 1+A_{1}\frac{\lambda
^{2}z^{2}}{2}+...+O\left( \lambda ^{6}z^{6}\right) \right]  \notag \\
&&\overset{z^{2}\ll 2\lambda ^{-2}}{\longrightarrow }\frac{L}{L^{2}-7}%
\lambda ^{2}z^{2}+\frac{1}{16}\left( 2A_{2}-A_{1}^{2}\right) \lambda
^{4}z^{4}+O\left( \lambda ^{6}z^{6}\right) .  \label{ms}
\end{eqnarray}%
(The second boundary condition can be read off from this expression.)
Equation (\ref{ms}) also implies that the solutions with $L=0,1$ turn
negative for $z\gtrsim 0$. At large $z$, on the other hand, the
inhomogeneity rises $\propto z^{4}$ and demands the modulus of $A_{S}$ to
grow as well. At some finite $z_{m}$ the nonlinearity $\propto \left(
e^{2A}-1\right) $ in Eq. (\ref{mde}) will therefore become too negative (for 
$A_{S}>0$ and $L<2$) to be counterbalanced by the derivative terms. To avoid
conflict with the increasingly positive inhomogeneity, the solution then
develops a singularity at $z_{m}$. As already mentioned, the sign of the
singularity depends on the second boundary condition. A negative singularity
occurs if the slope of $A_{S}$ near $z=0$ is smaller than a critical value
such that the argument of the logarithm in Eq. (\ref{amser}) eventually
reaches zero. This singularity is quantitatively reproduced by the series
solution (\ref{amser}) as long as $z_{m}$ lies inside its range of validity.

If the slope of $A_{S}$ at small $z$ exceeds the critical value, on the
other hand, the zero in the argument of the logarithm is avoided. Instead,
the argument stays positive with increasing $z$, passes through a minimum
and starts to increase until it reaches a positive pole singularity at
finite $z_{m}$. Singularities of this type lie outside the validity range of
the expansion (\ref{amser}) and imply that the non-conformal part of the
warp factor approaches a pole singularity as well. For $L=0$, e.g., it takes
the explicit form 
\begin{equation}
e^{2A_{S}\left( z\right) }\sim \frac{12}{\lambda ^{2}\left( z-z_{m}\right)
^{2}}.
\end{equation}

\paragraph{$L=2$:}

For $L=2$ the mass term in Eq. (\ref{mde}), and hence the strongest
nonlinearity, vanishes. The initial curvature $A_{S,L=2}^{\prime \prime
}\left( 0\right) =-4\lambda ^{2}/3$ is negative and again set by the
inhomogeneity (cf. Eq. (\ref{linL2})). Consequently, the solution can still
develop a negative singularity if its slope close to $z=0$ remains below a
critical value. For larger slopes the solutions turn positive and are 
nonsingular. Hence $L=2$ corresponds to the intermediate case in which the
solution space contains both singular solutions in which confinement
manifests itself by compactifying the fifth dimension (similar to the $L=0,1$
cases), and regular solutions analogous to those encountered for $L>2$ (see
below). Of course, the ensuing potential (\ref{vMconf}) is (up to $z_{m}$)
identical in both cases.

\paragraph{$L>2$:}

For all $L>2$ the main nonlinearity in Eq. (\ref{mde}) has a positive sign.
Moreover, at $z=0$ the solutions start out with positive curvature $%
A_{S}^{\prime \prime }\left( 0\right) =2L\lambda ^{2}/\left( L^{2}-7\right) $
(again dictated by the inhomogeneity, as can be seen from their linearized
counterparts (\ref{lL3})). In fact, the $A_{S,L>2}$ and the corresponding
warp factors remain positive and nonsingular at all $z$. Typical numerical
solutions for $A_{S}$ with $L=0,1,2$ and $3$ are displayed in Fig. \ref{am}.

\FIGURE{ \epsfig{figure=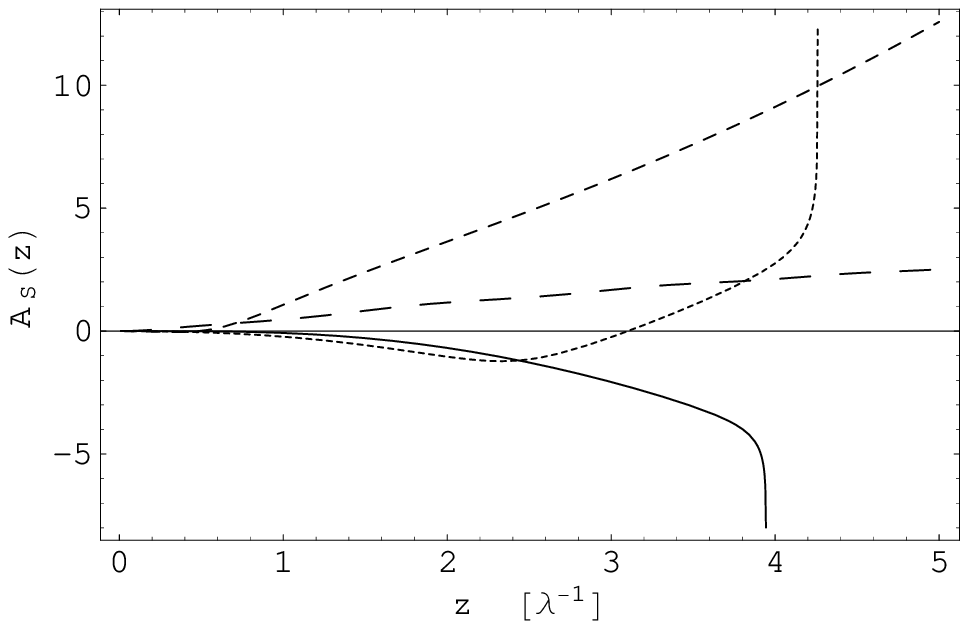}
\caption[dummy0]{
Typical solutions $A_{S}\left( z\right) $ for $L=0$ (full line, negative
sign of singularity selected), $L=1$ (dotted line, positive sign of
singularity selected), $L=2$ (short-dashed, absence of singularity
selected) and $L=3$ (long-dashed). Note that the dual eigenmodes have
significant support only for $z<\sqrt{2}\lambda ^{-1}$.}
\label{am}}

By construction, any background metric of the form (\ref{metric}), with $%
A_{S}$ a solution of Eq. (\ref{mde}), reproduces the potential (\ref{vMconf}%
) for all existing $z$. If the background space ends in $z$ direction at $%
z_{m}$, however, the potential (\ref{vMconf}) must end there, too. This does
not affect the (low-lying) spectrum as long as $z_{m}\gg \lambda ^{-1}$.
Hence for our purpose of maintaining the trajectory (\ref{mspec}) even when $%
L=0,1$ we choose the second boundary condition such that $z_{m}$ becomes as
large as needed\footnote{%
We have found numerical solutions of Eq. (\ref{mde}) with $z_{m}>6\lambda
^{-1}$ (for $L=0$).}. This selects the singularities of positive sign and
implies that the corresponding field modes\ automatically satisfy Dirichlet
(or Neumann) boundary conditions at $z_{m}$:%
\begin{equation}
f_{S,L=0,1}\left( z_{m}\right) =\left[ \lambda z_{m}e^{-A_{S}\left(
z_{m}\right) }\right] ^{3/2}\varphi _{L=0,1}\left( z_{m}\right) =0.
\label{fbv}
\end{equation}%
Even for radial excitations well beyond those currently experimentally
accessible, the $L=0,1$ part of the spectrum (\ref{mspec}) remains therefore
unaffected. (Note the Gaussian suppression of the eigenmodes (\ref{phi}) for 
$z_{m}^{2}\gg 2\lambda ^{-2}$.) We will elaborate on this issue in Sec. \ref%
{lsing}. (Recall for comparison that the much slower decay of the string
mode solutions in pure AdS$_{5}$ (Bessel functions) requires that boundary
conditions at the hard IR wall have to be imposed by hand. This strongly
modifies the spectrum - the masses become proportional to the zeros of
Bessel functions \cite{det05} - and generates the incorrect $M^{2}\propto
N^{2},L^{2}$ behavior which is typical for infinite square well or bag
potentials.)

\subsection{Vector meson sector}

The equation for the non-conformal warp factor in the spin-1 meson sector, 
\begin{equation}
-z^{2}A^{\prime \prime }+\frac{3}{2}z^{2}A^{\prime 2}-3zA^{\prime }+\frac{2}{%
3}\left( L^{2}-1\right) \left( e^{2A}-1\right) -\frac{2}{3}\lambda
^{2}z^{2}\left( \lambda ^{2}z^{2}+2L\right) =0,  \label{vde}
\end{equation}%
is obtained by setting the meson potential (\ref{vMconf}) equal to the
general-$A$ potential (\ref{vva}). This equation differs from its
counterpart (\ref{mde}) in the spin-0 sector only by the sign of the $%
A^{\prime \prime }$ term and by the replacement $\left( L^{2}-4\right)
\rightarrow \left( L^{2}-1\right) $ in the coefficient of the main
nonlinearity. The inhomogeneity and its $L$ dependence remain identical
since they  originate from the same bulk potential (\ref{vMconf}) in both
cases. Hence the analysis of the solution space is similar to that of Sec. %
\ref{sm} and will not be repeated in detail. Instead, we will just highlight
qualitative differences in the solution behavior and discuss the numerical
solutions of Eq. (\ref{vde}) and their implications.

The main differences between the solutions of Eqs. (\ref{mde}) and (\ref{vde}%
) can be rather directly traced to the two differences in the equations
mentioned above. The modified coefficient of the exponential term reflects
the fact that in the vector meson sector only $L=0$ corresponds to a
tachyonic mode while the $L=1$ mode is massless and all $L>1$ modes are
massive (cf. Eq. (\ref{m2v})). This is a consequence of the additional unit
of spin carried by the vector mesons. Hence the qualitative changes in the
solution behavior occur around $L=1$ (instead of at $L=2$). The sign flip of
the second-derivative term, furthermore, leads to changes in the sign of the
solutions $A_{V}\left( z\right) $ and in the development of singularities.
Part of these modifications are determined by the solution behavior at small 
$z$ which, as in the spin-0 case, can be obtained from the solutions to the
linearized version of Eq. (\ref{vde}) as derived in App. \ref{linode}.

The above qualitative expectations are corroborated by the numerical
solutions (again subject to the conformal boundary condition $A_{V}\left(
0\right) =0$). Their perhaps most important novel feature is that the second
boundary condition can be adapted to generate a common qualitative behavior
for all $L$ which is singularity-free. More specifically, every regular
solution turns negative\ towards larger $z$ where it decreases
monotonically. For $L>2$ the solutions start out with positive slope at $z=0$
(cf. Eq. (\ref{lvl2})) which implies that the decrease can set in only after
passing through a shallow, positive maximum at finite $z$. While the $L=0$
solution (associated with the tachyon mode) stays regular for all choices of
the second boundary condition, however, for $L>0$ one has the additional
possibility of solutions which remain positive towards larger $z$ and
develop a singularity at $z=z_{m}$ similar to those encountered in the
spin-0 sector for $L=0,1$. A set of typical solutions $A_{V}\left( z\right) $
for $L=0,...,3$ is plotted in Fig. \ref{av}. 
\FIGURE{ \epsfig{figure=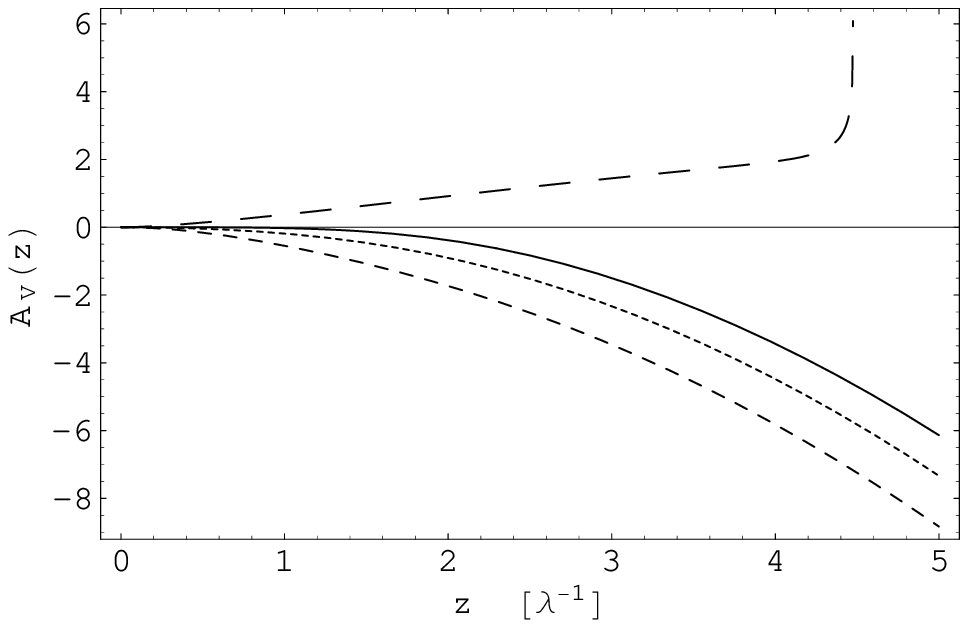}
\caption[dummy0]{
Typical solutions $A_{V}\left( z\right) $ for $L=0$ (full line), $L=1$ (dotted line), $L=2$ (short-dashed, absence of singularity
selected) and $L=3$ (long-dashed, singularity selected). Recall that the dual eigenmodes have
significant support only for $z<\sqrt{2}\lambda ^{-1}$.}
\label{av}}

To summarize, in the vector meson channel the dual manifestation of
confinement in terms of a dynamically compactified fifth dimension is not a
necessity but rather an option for the higher orbital excitations (with $L>1$%
). In contrast to the scalar sector, furthermore, the underlying
singularities are not associated with tachyonic modes.

We conclude this section by noting that the existence of simple gravity
duals which reproduce the linear trajectories (\ref{nltraj}), as constructed
above, provides additional support for the AdS/QCD program. Moreover, it
establishes the basis for calculating gauge theory correlation functions and
observables from the dual mode solutions (\ref{phi}) -- (\ref{psi-}) on the
basis of the AdS/CFT dictionary \cite{gub98}. Close to the AdS boundary, all
metrics found above have the same qualitative small-$z$ behavior with a
non-conformal warp factor of the form $\exp \left( cz^{2}\right) $ (except
for scalar mesons with $L=0$ and vector mesons with $L=2$), as determined by
the inhomogeneous terms in Eqs. (\ref{mde}), (\ref{bode}) and hence directly
induced by the underlying bulk potentials. This behavior suggests the
formation of a two-dimensional (nonlocal) gluon condensate \cite{gub01} and
indicates its relevance for linear confinement (cf. Sec. \ref{comp}). (For $%
L=0$ mesons the standard four-dimensional gluon condensate seems to
dominate, on the other hand, which should be compared to the operator
product expansion with renormalon-type corrections for the corresponding
QCD\ correlators.) At large $L$, furthermore, the leading $z$ dependence of
the mesonic and baryonic warp factors becomes identical. Finally, an IR
cutoff dual to $\Lambda _{\text{QCD}}$ for the fifth dimension emerges as a
necessary requirement for weakly orbitally (i.e. $L=0,1$) excited scalar
mesons and as a possibility for $L>1$ vector mesons. In our framework this
compactification of the fifth dimension occurs dynamically.

\section{Discussion of the resulting holographic duals \label{metdisc}}

In the following section we elaborate on several important properties of the
IR-deformed gravity backgrounds found above and discuss their physical
significance.

\subsection{Singularities in the meson sector and linear confinement\label%
{lsing}}

In several recent holographic QCD models, the asymptotically free and hence
almost conformal region (with at most weakly deformed AdS metric) is assumed
to extend down to energies $z^{-1}$ of the order of the QCD scale. It gets
broken only in the infrared, by a rather abrupt onset of nonperturbative
effects including condensates and confinement.

This scenario is indeed borne out dynamically by our results especially for
the lowest (i.e. $L=0,1$) orbital spin-0 meson excitations. These are
exactly the meson modes accessible in weakly curved dual supergravity
backgrounds. As shown above, for them to lie on linear trajectories requires
a singular metric of the type \cite{mal98,kin00} emerging in holographic
duals of confining theories. In these cases the onset of confinement indeed
occurs rather sudden, although not as sudden as in the extreme hard wall
case (\ref{hw}). We have already traced the origin of these singularities to
the tachyonic nature of the corresponding dual modes\footnote{%
It may be tempting to speculate about potential relations between these
tachyon-induced singularities and closed string tachyon condensation \cite%
{yan05} in the bulk, which is expected to be dual to confinement on the
gauge theory side (see also Refs. \cite{gre00,csa06}).}. In the vector meson
sector, in contrast, singular metrics are possible but not mandatory
(depending on the choice of a boundary condition) for orbitally excited
resonances, and they are not tachyon-induced.

Physically, it seems that no ($L=0$)\ or only a small ($L=1$) centrifugal
barrier allows the corresponding spin-0 meson states to probe different
aspects of the IR region and hence to feel a more sudden onset of
confinement, in particular at large $N$. The location of the singularities,
however, is set by the inverse mass scale in the second boundary condition
and can therefore be put at $z_{m}\gg \lambda ^{-1}$ large enough not to
significantly alter the low-lying part of the spectra. Furthermore,
semiclassical arguments indicate that highly excited hadrons generally
become larger and therefore should be able to explore more of the IR region
as well. This may explain why the vector meson metric can become singular
for $L>1$.

The possibility to set $z_{m}\gg \lambda ^{-1}$ implies, in particular, that
one could choose extensions of the metric into the region $z\in $ $\left[
z_{m}-\varepsilon ,\infty \right] $ which yield the same low-lying spectra
and wave functions (cf. Eq. (\ref{fbv})) without any singularities. In the
intermediate case of spin-0 mesons with $L=0$ the choice between singular
and regular metrics exists even inside the solution space of Eq. (\ref{mde}%
). For scalar meson excitations with $L>2,$ vector mesons under suitable
boundary conditions and all baryons, finally, confinement does not manifest
itself in metric singularities at all.

Nevertheless, all higher meson and all baryon excitations are found to lie
on the linear trajectories\footnote{%
The holographic models of Refs. \cite{kar06,and06} also realize linear meson
trajectories with a non-singular metric, but they contain an additional
(regular) dilaton field.} (\ref{mspec}) and (\ref{bspec}). The combination
of these results may reflect the fact that the wave functions of light
hadrons seem to be rather weakly affected by the linear confinement force.
Indeed, several successful models for low-lying hadrons (e.g. models of
Skyrme-type \cite{zah86} and the instanton-based chiral quark model \cite%
{dia88}) do not implement confinement at all since the mean separation among
colored constituents appears to be too small for confinement effects to
become relevant.

As noted in Ref. \cite{det05}, the abrupt hard wall singularity of the
metric (\ref{hw}) - and the IR boundary conditions it requires - resemble
those of an MIT bag model with a sharp surface \cite{cho75}. In our case the
singularities develop more gradually. The analogous bags therefore have
smooth transition regions as they emerge dynamically in soliton bag models
of Lee-Friedberg \cite{lee77} or color-dielectric \cite{nie82} types. Such
soliton bags become confining by means of a space-dependent color dielectric
function which induces singular couplings to vacuum fields \cite{for86}.
This suggests that the IR deformations of the dual gravity background found
above encode information on the color-dielectric QCD vacuum structure.

String breaking due to light quark production is expected to stop the linear
rise of the QCD confinement potential at large distances and hence to bend
the linear hadron trajectories\ at sufficiently high excitation levels.
Although such effects are not yet visible in the experimentally accessible
part of the hadron spectrum, string breaking has recently been confirmed on
the lattice \cite{bal05}. In our model, however, the linearity of the
spectra (\ref{mspec}), (\ref{bspec}) continues up to arbitrarily high
excitation quanta (except for spin-0 mesons in $L=0,1$ states with $%
N\rightarrow \infty $, due to the finite $z$ effects discussed above). This
indicates that string breaking effects are absent in our holographic dual,
as expected in the large-$N_{c}$ limit (where $N_{c}$ is the number of
colors) or the associated weak coupling approximation on the gravity side.

A simple way to account for string breaking effects by hand would be to
modify the solutions $A$ at $z>z_{m}-\varepsilon $ such that the potential
levels off. As already alluded to, for $\left( \lambda z_{m}\right) ^{2}\gg
1 $ any reasonably smooth deformation of $A$ at large $z$ may in fact be
implemented with practically no impact on the low-lying part of
wavefunctions and spectra. In this way one could for example remove the
singularities altogether. Not surprisingly, this also implies that the
generation of the linear trajectories (\ref{nltraj}) does not fully
constrain the IR behavior of the gravity background. Together with the
second boundary condition in the meson sector, the remaining freedom could
be used to implement a more comprehensive set of QCD observables.

\subsection{$L$ dependence \label{ldep}}

The IR deformations of the AdS$_{5}$ metric obtained from the solutions of
Eqs. (\ref{bode}), (\ref{mde}) and (\ref{vde}) are necessarily $L$
dependent. This $L$ dependence enters through the potentials (\ref{vMconf}),
(\ref{vBconf}) which give rise to the inhomogeneities of the differential
equations for $A_{S,V,B}$, and more universally through their counterparts (%
\ref{vma}) -- (\ref{vba}) in the IR deformed background. Its ultimate source
is therefore the $L$-dependent twist dimension of the considered hadron
interpolators, imposed via Eqs. (\ref{m2m}) -- (\ref{mb}), and its main
effects are independent of the specific choice for the heuristic potentials
or the replacement rule (\ref{rrule}). A somewhat analogous hadron
dependence of a dual background has been found in Ref. \cite{hir06}, where
vector and axial vector mesons feel a different metric, and would enter
several other holographic models if observables in the whole hadron spectrum
were to be reproduced.

A natural source for the $L$ dependence arises in our approach from the
identification of orbital hadron excitations as stringy quantum fluctuations
about the AdS background \cite{det05,bro04}. Indeed, such $L$ dependent
fluctuations may deform the AdS background metric in an $L$ dependent
fashion. In the simpler case of two-dimensional quantum gravity,
fluctuation-induced deformations of a background metric (essentially AdS$%
_{2} $) were recently found explicitly \cite{amb06}. In our case, the
back-reaction of the metric to a fluctuation dual to a given orbital
excitation could conceivably lead to analogous, $L$ dependent deformations,
as found in our solutions $A_{S,V,B}\left( z\right) $. Although the
different orbital excitations \emph{feel} a different total metric, however,
the overall conformal symmetry breaking scale $\lambda $ remains (almost)
hadron independent. This is a consequence of its relation (\ref{slope}) to
the almost universal slope of all hadron trajectories, which we will discuss
quantitatively in Sec. \ref{phen}.

The identification of orbital excitations as duals of metric fluctuations
sets their holographic origin apart from that of the radial (i.e. $N$)
excitations. This becomes manifest in the fact that our metric acquires no $%
N $ dependence while even more sophisticated IR deformations would
necessarily be $L$ dependent since each orbitally excited hadron state is
created by a different interpolator. Via back-reactions similar to those
considered in Refs. \cite{csa06,sho07} this $L$ dependence may carry over to
additional background fields of stringy origin as well. The relative
weakness of the warp factors' $L$ dependence in the physically dominant
region $z<\sqrt{2}\lambda ^{-1}$ and for larger $L$ (where it becomes
identical in the meson and baryon sectors) may be another indication for
their fluctuation-induced origin.

Additional support for the above interpretation of the angular-momentum
dependence in our background arises from an observation in Ref. \cite{kar06}%
. The latter asserts that for on-shell gauge theory properties which are
described by the quadratic part of the dual string action, like the
trajectories (\ref{nltraj}), the effect of higher derivative terms
(including those related to orbital angular momentum operators, cf. Sec. \ref%
{ads}) can be essentially reproduced by the standard quadratic terms - to
which we restrict ourselves here - in a modified gravity background. The
modifications will depend, in particular, on the hadronic angular momentum
carried by the dual modes, as manifested in our case in the solutions $%
A_{S,V,B}$.

Pursuing the above line of reasoning farther, we recall that in the
holographic model based on the hard-wall metric (\ref{hw}) quantum
fluctuations corresponding to orbital excitations are (at least in the
conformal regime at small $z$) represented by an $L$ dependent effective
mass for the bulk string modes. Our derivation of the dual gravity
backgrounds suggests a generalization of this interpretation. By allowing
the effective masses to change with resolution $r=R^{2}/z$ outside of the
conformal regime (i.e. for $z>0$, as via the replacement (\ref{rrule})) one
may be describing the quantum fluctuations of the metric and their
potentially deforming back-reaction in more detail, and hence obtain a more
accurate description of IR properties (including the linear hadron
trajectories (\ref{nltraj})) on the gauge theory side.

\subsection{Background field content and underlying dynamics}

The construction of our background geometry raises the question how it may
be related to an underlying string theory. One could start to gain insight
into this matter by examining, for example, whether the differential
equations (\ref{bode}), (\ref{mde}) and (\ref{vde}) for the metric can be at
least approximately cast into the form of (potentially higher-dimensional)
Einstein equations. Such issues are beyond the scope of the present paper,
however, where we focus on general principles, symmetries and experimental
data to constrain the holographic dual in bottom-up fashion but leave the
underlying dynamics at least \emph{a priori} undetermined. In the remainder
of this section we will therefore only mention a few qualitative aspects of
the dynamics expected to govern the dual background and summarize the
rationale for restricting our construction to the metric. 

As stated in the introduction, the present knowledge of string theory in
strongly curved spacetimes does not provide \emph{ab initio} insight into
the background field composition and dynamics of the QCD dual. (Guidance
from the supergravity approximation, in particular, is limited even at large 
$N_{c}$: it would predict a mass gap far smaller than the string tension
beyond which linear trajectories can emerge, for example, in conflict with
QCD phenomenology \cite{kar06}.) Nevertheless, several approximate, QCD-like
duals, partly including fundamental flavor as mentioned in Secs. \ref{intro}
and \ref{ads}, have recently been obtained from brane constructions in which
the metric and other background fields are solutions of classical field
equations. Other dual models were derived by solving simple, string-inspired
scalar and Einstein equations \cite{csa06,sho07} and used to get an idea of
the impact of QCD condensates on the background. These models show, in
particular, that such back-reactions can generate confining IR cutoffs in
the fifth dimension similar to those encountered in our approach. 

Our exclusive reliance on the dual background metric, finally, was guided by
the dimensional power-counting argument of Sec. \ref{metderiv}, the
possibility to represent the impact of additional background fields on the
resonance masses by deformations of the metric (cf. Sec. \ref{ldep}), and by
an Occam-razor type preference for the minimal and hence most efficient
background required to reach our objectives. Nevertheless, experience from
brane models as discussed above suggests that a more comprehensive and
detailed approximation to the QCD dual should contain additional background
fields. As pointed out in Ref. \cite{kar06}, for example, one may expect a
background including tachyon and dilaton fields if the dual confinement
mechanism has its origin in closed-string tachyon condensation (see also
Ref. \cite{csa06}).

\subsection{Comparison with other confining holographic models \label{comp}}

In the following we will briefly compare our holographic model to a few
related approaches which also contain dual representations of linear
confinement and linear trajectories in the hadron spectrum. As already
mentioned, it turns out to be a nontrivial task to reproduce linear
trajectories with approximately universal slopes not only in the meson but
also in the baryon channels. In fact, our approach seems to be the first
which accomplishes this. Hence our comparisons below have to remain
restricted to the meson sector.

Linear ``meson'' trajectories in more or less QCD-like gauge theories were
e.g. found in Refs. \cite{kru05,and206}. A recent implementation of linear
trajectories for both radial and spin excitations of the rho meson into the
AdS/QCD framework \cite{kar06} induces conformal symmetry breaking mainly by
a dilaton background field. In the simplest case a dilaton of the form $\Phi
\left( z\right) \propto z^{2}$ is added to the pure AdS metric and hence is
exclusively responsible for the non-conformal IR behavior. (Essentially the
same term was argued to arise from a dual magnetic condensate (which plays
the role of a Higgs field) in the partition function of a QCD instanton
ensemble when promoted into the bulk \cite{shu06}.) Although this approach
works well in the (vector) meson sector, we have already noted that such
dilaton effects do not manifest themselves in the baryon spectrum since they
can be absorbed into the eigenmodes and leave the AdS/CFT boundary condition
unchanged. An interesting observation of Ref. \cite{kar06} is that the
exponent of the warp factor should not contain contributions growing as $%
z^{2}$ for $z\rightarrow \infty $, in order to have spin-independent slopes
of the radial rho meson excitation trajectories. Our solutions $A_{S,V,B}$
grow logarithmically with $z$ for large\footnote{%
except if the large-$z$ region is cut off by a singularity, of course} $z$
and are therefore consistent with this condition (while the model of Refs. %
\cite{and06,and206} is not, see below).

In other recent work, scalar bulk fields dual to the gluon and bilinear
quark condensate operators have been shown to generate - by their
back-reaction on the gravity background - a confining restriction of the
metric to a deformed\ AdS slice \cite{csa06,sho07}. Indications for the
potential role of the two-dimensional QCD condensate \cite{gub01} in the
confinement mechanism arise in these frameworks as well. As already
mentioned, the non-conformal behavior of our metric towards small $z$,
induced by the confinement-generating modifications of the string mode
potentials, may similarly be a reflection of the non-local dimension-two
condensate. This suggest a more general analysis of the information on QCD
condensates and more specific IR degrees of freedom (e.g. topological ones
related to instantons, monopoles or center vortices) \cite{for06} which is
encoded in our background metric.

The qualitative small-$z$ behavior $\exp \left( cz^{2}\right) $ of our
solutions for the non-conformal warp factors is identical to that proposed
in Refs. \cite{and06,and206} where it has been shown to embody, together
with a constant dilaton, a linearly growing heavy-quark potential and a
linear (mesonic) Regge trajectory \cite{and06,and206}. The
value of $c$ was estimated to be $c\sim -0.9$ GeV$^{2}$ (in a metric with
Lorentzian signature) which is in the same ballpark as ours for intermediate 
$L$. Indeed, anticipating the relation (\ref{Lam}) between $\lambda $ and $%
\Lambda _{\mathrm{QCD}}$ to be established in Sec. \ref{phen}, our warp
factor implies e.g. $c_{S,L=2}\simeq -0.7$ GeV$^{2}$. Our values for $c_{B}$
in the baryon sector are positive, however, which may suggest some
differences in the dual confinement mechanisms for vector mesons and baryons.

\section{Phenomenological implications \label{phen}}

In the following section we proceed to the quantitative analysis of our
holographic dual and confront the predicted mass spectra (\ref{mspec}), (\ref%
{bspec}) for the light hadrons with experimental data. Recent reviews of
excited hadrons, their symmetry structure, parity doubling etc. can be found
in Refs. \cite{jaf06}.

We start by determining the conformal symmetry breaking scale $\lambda $
from data for the slope $W=4\lambda ^{2}$ of the trajectories (\ref{nltraj}%
). Fits to the experimental meson spectra~yield $W=(1.25\pm 0.15)\ \mathrm{%
GeV}^{2}$~\cite{ani00} and $W=(1.14\pm 0.013)\ \mathrm{GeV}^{2}$ \cite{bug04}%
. These values allow for an immediate check of our relation (\ref{mwmrel})
which predicts the rho meson mass as a function of its trajectory slope,
i.e. 
\begin{equation}
M_{\rho }=\sqrt{\frac{W}{2}}.
\end{equation}%
The above empirical results for $W$ imply $M_{\rho }=0.79\ \mathrm{GeV}$ or $%
M_{\rho }=0.76\ \mathrm{GeV}$, respectively, which are both consistent with
the experimental value $M_{\rho }=0.7755\pm 0.0004$ GeV~\cite{pdg}. Since
the latter is close to the mean of the slope fit results, we choose the
experimental rho mass to set the scale of the deformed gravity background,
i.e. 
\begin{equation}
\lambda =\sqrt{\frac{W}{4}}=\frac{M_{\rho }}{\sqrt{2}}=0.55\ \mathrm{GeV}.
\label{lamrho}
\end{equation}%
The corresponding value $W=4\lambda ^{2}=\allowbreak 1.21$ $\mathrm{GeV}^{2}$
fixes the slope of our linear meson trajectory which is compared to the
experimental meson resonance spectrum (for quark-antiquark states) in Fig.~%
\ref{fig1}. 
\FIGURE{ \epsfig{figure=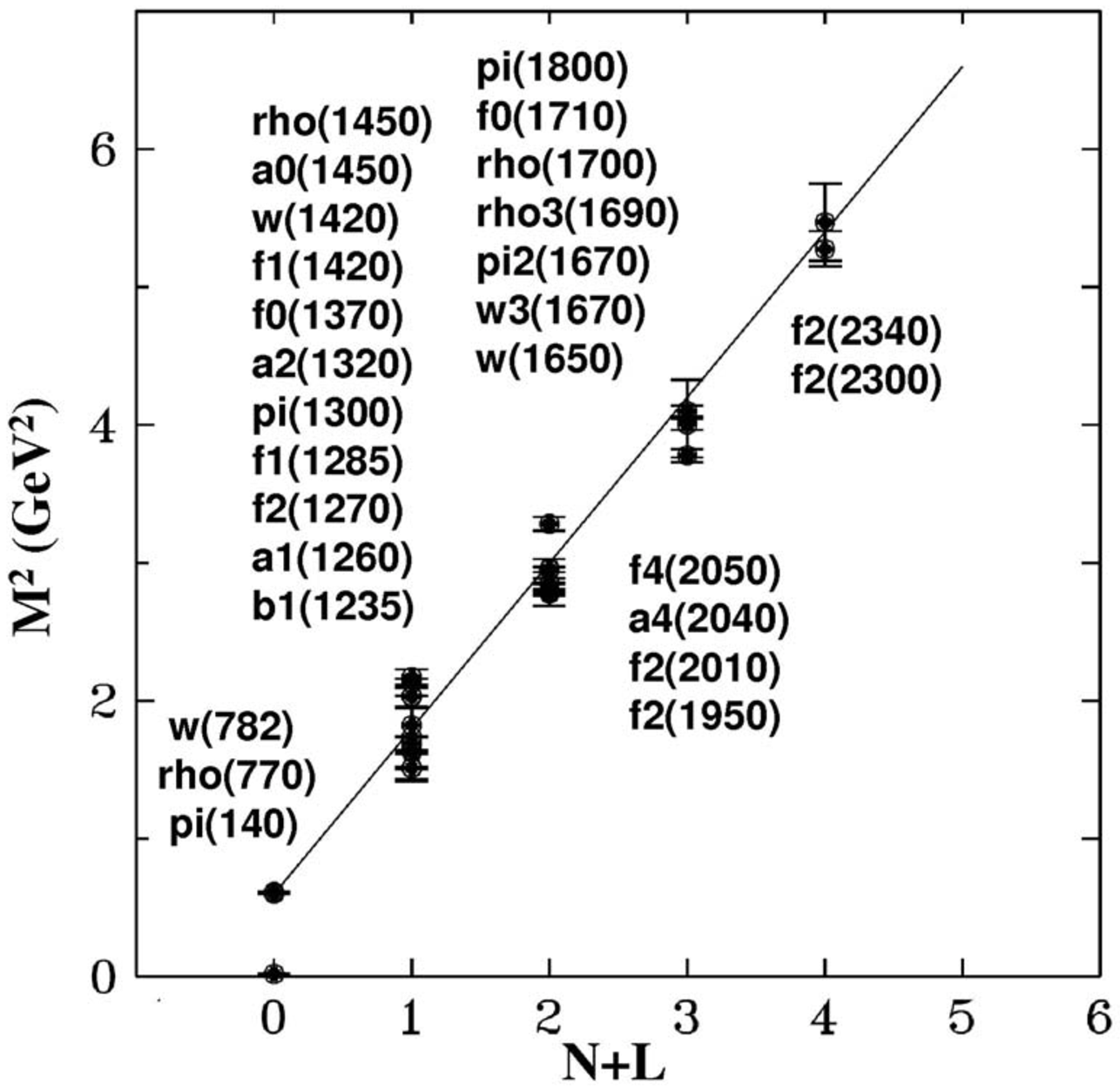}
\caption[dummy0]{Experimental meson mass spectrum from Ref. \protect\cite{pdg} 
and the predicted trajectory for $W=2M_\protect\protect\rho^2
\simeq 1.01$ GeV$^2$.}
\label{fig1}}

The clustering of radial and orbital excitations is clearly visible, and
even the highest radial excitations $f_{2}(2300)$ and $f_{2}(2340)$ lie
squarely on the linear trajectory. The pion ground state, set apart by its
approximate Goldstone boson nature, does not fit into the overall pattern
predicted by the dual string modes. This problem is expected. It was already
encountered in Ref. \cite{det05} and is caused by the lack of chiral
symmetry and its breaking in our approximate holographic dual. For the same
reasons, qualitative models for the light-front square mass operator in the
mesonic valence quark sector~\cite{FPZ02} (which reproduce the radial
excitation trajectory) need a strong additional short-range attraction in
the spin-$0$ channel to reproduce the $\pi $-$\rho $ mass splitting (for $%
L=0 $). Since the dual string modes are related to the valence components of
the light-front wave function (with $z$ playing the role of a relative
coordinate)~\cite{bro06}, the impact of such interactions is accessible in
our approach.

The empirical slope of the $\Delta $ trajectory (including the nucleon
resonances in the $^{4}8$ representation of SU(4)) is $W=(1.081\pm 0.035)\ 
\mathrm{GeV}^{2}$ \cite{kle02}. As in the meson sector, our relation (\ref%
{mwbrel}) turns this value into a prediction for the $\Delta $ ground state
mass, 
\begin{equation}
M_{\Delta }=\sqrt{\frac{3W}{2}}=1.27\ \mathrm{GeV,}
\end{equation}%
which compares well with the experimental value $M_{\Delta }=1.232$ GeV.
This suggests to use the experimental mass of the $\Delta $ isobar for an
alternative determination of the scale 
\begin{equation}
\lambda =\frac{M_{\Delta }}{\sqrt{6}}=0.50\ \mathrm{GeV}  \label{lamdel}
\end{equation}%
in the baryon sector. The value (\ref{lamdel}) differs by less than 10\%
from that based on the experimental rho mass, Eq. (\ref{lamrho}). This
confirms the approximate universality of $\lambda $ and of the associated
slopes $W=4\lambda ^{2}$ in the meson and baryon sectors.

The resulting $\Delta $ isobar trajectory, together with the empirical
square masses of the radial and orbital resonances, is shown in Fig.~\ref%
{fig2}. 
\FIGURE{\epsfig{figure=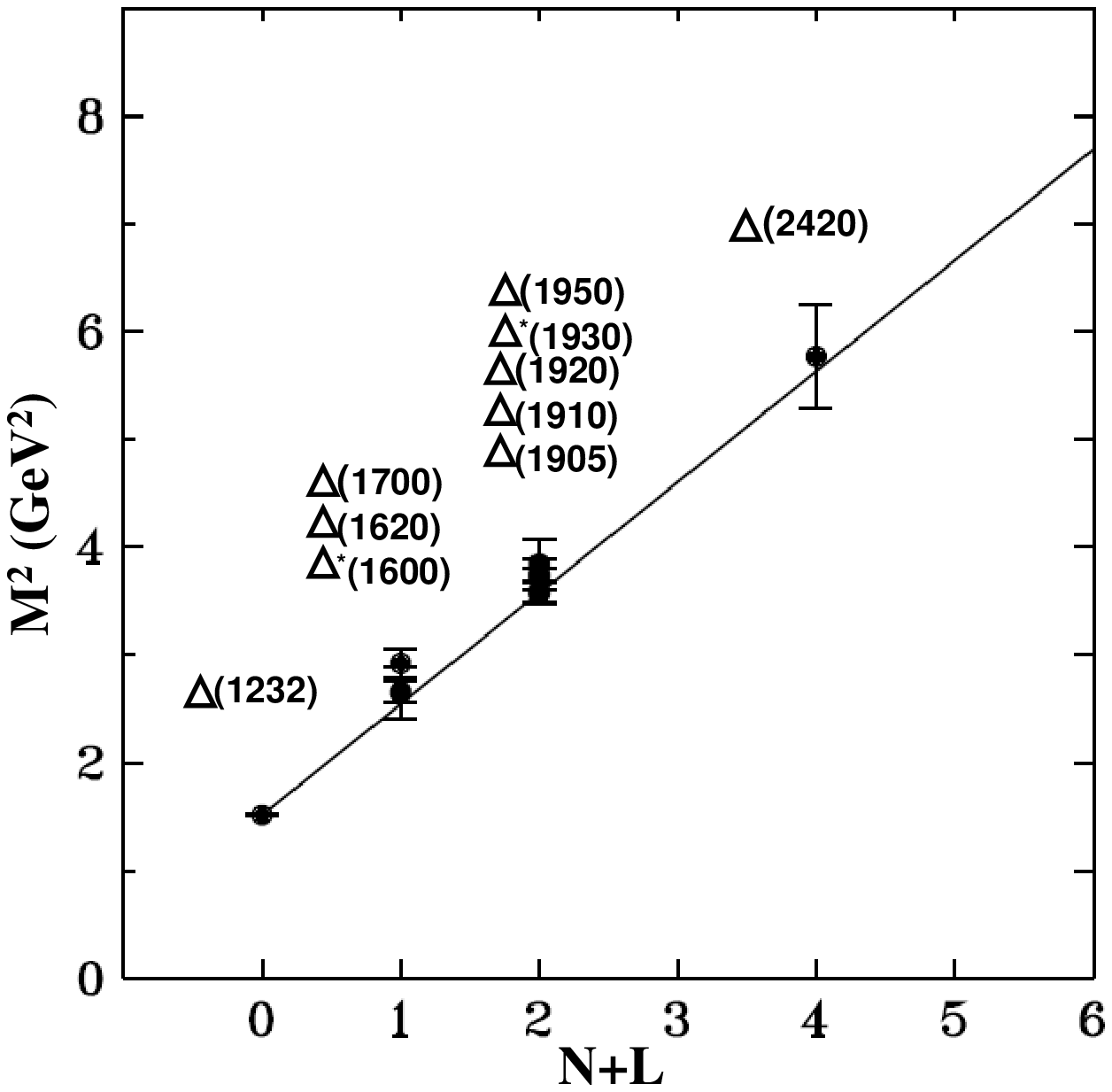}
\caption[dummy0]{Experimental Delta isobar mass spectrum from Ref. 
\protect\cite{pdg} and the predicted trajectory for $S=3/2$ (in the $^4$8
representation of SU(4)) with $W=2M_\Delta^2/3 \simeq 1.01$ GeV$^2$. }
\label{fig2}} The first radial excitations, $\Delta (1600)\frac{3}{2}^{+}$
(with $L=0$) and $\Delta (1930)\frac{5}{2}^{-}$ (with $L=1$) are degenerate
with states carrying one or two units of angular momentum, respectively. The
parity doublets $\Delta (1600)\frac{3}{2}^{+}$, $\Delta (1700)\frac{3}{2}%
^{-} $ and $\Delta (1905)\frac{5}{2}^{+}$, $\Delta (1930)\frac{5}{2}^{-}$,
furthermore, are states which differ by one radial and one orbital
excitation quantum such that $N+L$ is preserved.

Finally, we turn to the nucleon and its excitations. As noted in Ref. \cite%
{kle02}, the trajectory of the nucleon resonances in the $^{2}8$
representation of SU(4) (including the nucleon itself) lies below that for
the $^{4}8$ representation. This behavior can be accommodated by our
covariant\footnote{%
Note that the identification of the not separately Lorentz-invariant orbital
angular momentum $L$ requires us to select a particular frame, which
underlies the interpretation of our interpolators (cf. Sec. \ref{ads}).}
framework with a somewhat smaller value of $\lambda =0.47\ \mathrm{GeV}$
which may e.g. be due to hyperfine interactions. The resulting trajectory
indeed fits the nucleon resonances in the $^{2}8$ representation well and is
shown as a solid line in Fig.~\ref{fig3}. 
\FIGURE{\epsfig{figure=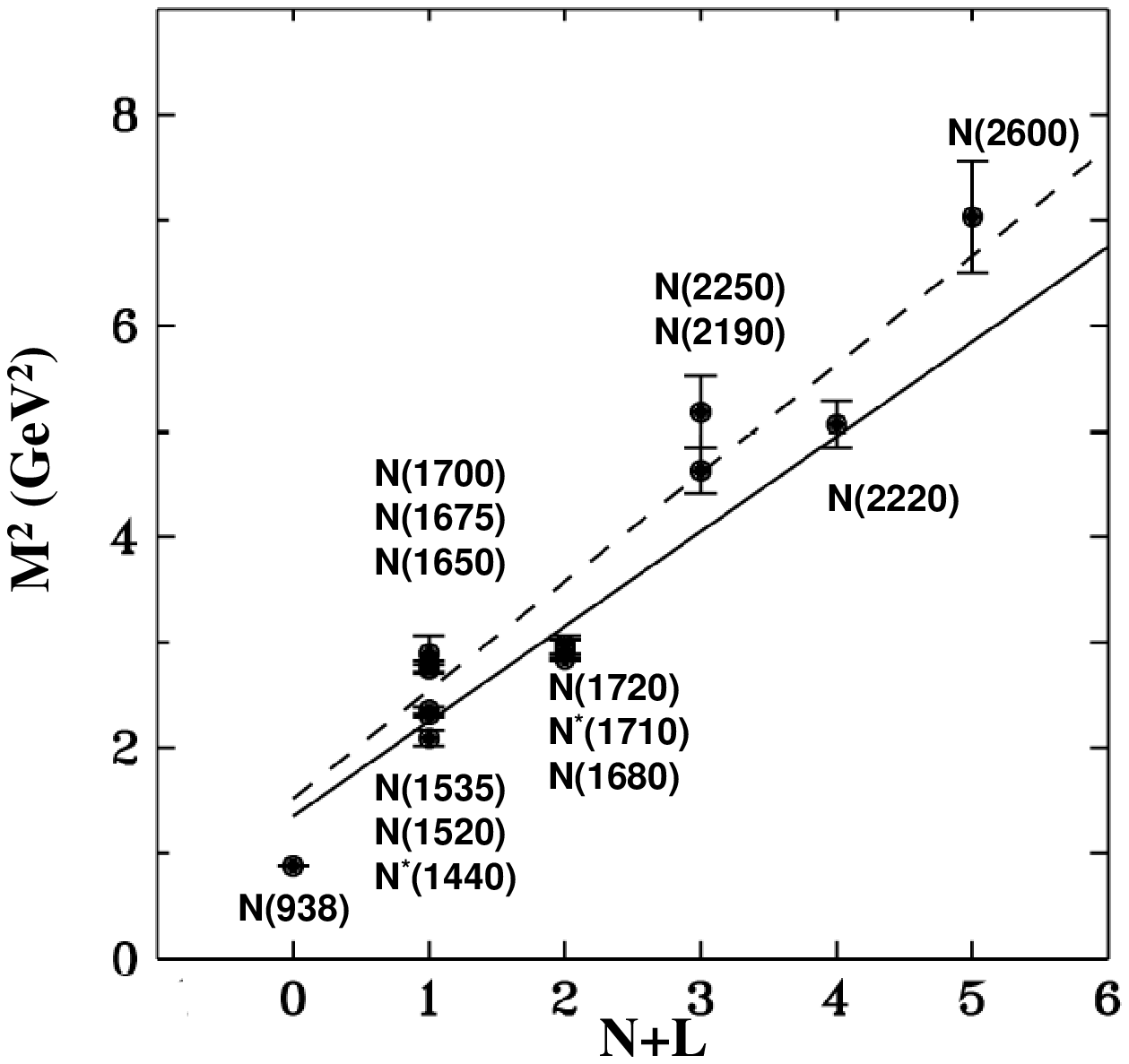}
\caption[dummy0]{Experimental nucleon mass spectrum from Ref. 
\protect\cite{pdg} and the predicted trajectories for
$S=1/2$ with $W \simeq 0.9$ GeV$^2$ (solid line) 
and for $S=3/2$ with $W=2M_\Delta^2/3 \simeq 1.01$ GeV$^2$ (dashed line).}
\label{fig3}} The experimental value of the nucleon mass lies below the
value $M_{N}=1.16\ \mathrm{GeV}$ on the trajectory, however. This may again
be related to chiral symmetry breaking effects and implies, in any case,
that the resulting nucleon delta splitting vanishes. Since the latter is an $%
O\left( 1/N_{c}\right) $ effect (where $N_{c}$ is the number of colors),
this result is consistent with the large-$N_{c}$ limit, i.e. the weak string
coupling limit which underlies all known top-down holographic duals. The
trajectory of Fig.~\ref{fig2} is included as a dashed line in Fig.~\ref{fig3}
and seen to fit the nucleon resonances in the $^{4}8$ representation, as
anticipated.

Our holographic results for the at present experimentally accessible
orbitally excited nucleon states are generally close to those of Ref.~\cite%
{det05} (which were based on the hard IR wall metric (\ref{hw})) because at
moderate $N+L$ the difference between linear and quadratic trajectories is
rather small. In addition, we predict the radially excited states, i.e. the
Roper resonance $N(1440)$ and the second radially excitation $N(1710)$,
which is almost degenerate with the $L=2$ states $N(1680)$ and $N(1720)$.
Our results for nucleons with internal spin 3/2 lie approximately on the $%
\Delta $ trajectory (dashed line), as already mentioned. The parity doubling
of baryon states with fixed total spin, differing by one unit of angular
momentum and one internal spin or radial excitation quantum, emerges
naturally in our approach.

The slope $W$ of our trajectories is related to the QCD scale. As a
consequence of confinement, the Gaussian suppression factor $\exp \left[
-\left( W/8\right) z^{2}\right] $ prevents the dual string modes (\ref{phi})
- (\ref{psi-}) from extending significantly beyond distances $z_{m}\sim 
\sqrt{8/W}$ into the fifth dimension. Similar confinement effects are often
modelled by the hard IR wall metric (\ref{hw}) with $z_{m}\sim \Lambda _{%
\mathrm{QCD}}^{-1}$, as discussed in Sec. \ref{metderiv}. Hence we
approximately identify 
\begin{equation}
\Lambda _{\mathrm{QCD}}\simeq \sqrt{\frac{W}{8}}=\frac{\lambda }{\sqrt{2}}%
\simeq 0.35~\text{GeV.}  \label{Lam}
\end{equation}%
The numerical estimate in Eq. (\ref{Lam}) is based on the phenomenological
slopes of about 1~GeV$^{2}$ and indeed close to the empirical value $\Lambda
_{\mathrm{QCD}}$ $\simeq 0.33$~GeV (at hadronic scales with three active
flavors) \cite{pdg}. Finally, we recall that the string tension resulting
from the semiclassical treatment of simple, relativistically rotating string
models \cite{shi05}, i.e. 
\begin{equation}
\sigma =\frac{W}{2\pi }\simeq 0.88\text{ }\frac{\text{GeV}}{\text{fm}},
\label{st}
\end{equation}%
is consistent with standard values as well.

\section{Summary and conclusions \label{sum}}

We have shown how a salient empirical pattern in the light hadron spectrum,
namely the combination of both radial and orbital excitations into linear
trajectories of approximately universal slope, can be reproduced with good
accuracy by a rather minimal version of holographic QCD. Our approximate
holographic dual relies exclusively on IR\ deformations of the AdS metric,
governed by one free mass scale proportional to $\Lambda _{\text{QCD}}$, and
generates the mass gap expected from confining gauge theories. Moreover, it
provides the first example of a gravity dual which is able to reproduce
linear trajectories in the baryon sector as well.

The resulting light hadron spectra are in good overall agreement with the
available\ experimental data for both meson and baryon masses. Discrepancies
between the radial and orbital resonance masses in the hard-wall model are
resolved and the experimentally established, approximately\ universal slope
of the trajectories emerges naturally. Moreover, new relations between the $%
\rho $ meson and $\Delta $ isobar ground state masses and the slopes of
their respective trajectories are predicted. Since the linearity of all
trajectories extends to the lightest masses, however, they fail to reproduce
the physical pion and nucleon ground states. This is not unexpected because
our approximate gravity dual in its present form lacks information on chiral
symmetry and residual interactions responsible e.g. for hyperfine splittings.

Our holographic background was derived by reconstructing the dual mode
dynamics from spectral properties on the gauge theory side. The underlying
strategy may be useful for other applications as well. It consists of first
finding the modifications of the AdS string mode potentials which generate a
desired gauge theory result, and to subsequently construct the corresponding
fields on the gravity side by equating the potentials induced by a general
background to their heuristic counterparts. As long as a dual background
exists, its derivation is then reduced to solving the resulting differential
equations.

Above we have constructed the holographic duals for radial and orbital
hadron trajectories in a minimal way, i.e. by a non-conformal warp factor. A
remarkable \textit{a posteriori} justification for this restriction is its
sufficiency. More complex IR deformations and further bulk fields would also
introduce new parameters to be fixed by QCD\ phenomenology and hence lessen
the predictivity of the dual description. In the baryon sector we were able
to derive the resulting IR deformations of the metric analytically. The
lowest orbital excitations of the spin-0 mesons, and depending on a boundary
condition also the higher orbital vector meson excitations, encounter a
singular metric. This is a dual confinement signature and results in a
dynamical compactification of the fifth dimension, hence directly linking
linear trajectories (for not too high excitation levels) to linear quark
confinement.

Despite the advantages of the minimal description, however, experience from
supergravity and brane models suggests that a more comprehensive holographic
dual may require a more general form of the metric and additional background
fields. The prospect of deriving those by our method deserves further
investigation. Among the potentially useful extensions and applications we
mention the implementation of quark flavor and spontaneously broken chiral
symmetry as well as the calculation of condensates, heavy-quark potentials
and light-front wave functions.

\begin{acknowledgments}
We thank Stan Brodsky for emphasizing the uses of the AdS/CFT correspondence 
in strong interaction physics and are grateful to him and Guy~de 
T\'{e}ramond for constructive comments on an early draft of the manuscript. We 
acknowledge partial financial support by the Brazilian and German  funding 
agencies CAPES/DAAD, Funda\c{c}\~{a}o de Amparo a Pesquisa do Estado de 
S\~{a}o Paulo (FAPESP) and Conselho Nacional de Desenvolvimento 
Cient\'{\i}fico e Tecnol\'{o}gico (CNPq).
\end{acknowledgments}

\appendix

\section{Solutions of the linearized differential equations for the mesonic
warp factor \label{linode}}

In this appendix we solve the linearized versions of the differential
equations (\ref{mde}) and (\ref{vde}) for the non-conformal mesonic warp
factors. The results will be useful for our analysis of the solution
behavior of the full, nonlinear equations in Sec. \ref{metderiv}. It will
also be instructive to understand the differences in the solution behavior
induced by tachyonic, massless and massive modes in this simplified setting,
although the solutions of the linearized equations do not develop finite-$z$
singularities.

The linearization of the differential equation (\ref{mde}) for the spin-0
meson sector leads to the (still inhomogeneous) equation%
\begin{equation}
z^{2}A^{\prime \prime }-3zA^{\prime }+\frac{4}{3}\left( L^{2}-4\right) A-%
\frac{2}{3}\lambda ^{2}z^{2}\left( \lambda ^{2}z^{2}+2L\right) =0
\label{A0lin}
\end{equation}%
whose solutions provide approximations to those of the full equation in the
regions where $A\ll 1$. As a consequence of the conformal boundary condition 
$A\left( 0\right) =0$, which turns out to be an automatic property of all
solutions which stay finite at $z=0$, this condition should hold in
particular in the UV, i.e. for $z$ close to zero.

The full solution space of the linear equation (\ref{A0lin}) can be
constructed by Frobenius expansion techniques (for the homogeneous part) and
by guessing special solutions of the full, inhomogeneous equation or by
deriving them with the help of Green function methods. Either way, the
general solution for $L=0,1$ (associated with the tachyonic string modes in
the bulk) is found to be 
\begin{equation}
\bar{A}_{S,L=0,1}\left( z\right) =\frac{L\lambda ^{2}z^{2}}{L^{2}-7}+\frac{%
\lambda ^{4}z^{4}}{2L^{2}-8}+c_{1}\left( \lambda z\right) ^{2+2\sqrt{\frac{%
7-L^{2}}{3}}}+c_{2}\left( \lambda z\right) ^{2-2\sqrt{\frac{7-L^{2}}{3}}}.
\label{linL01}
\end{equation}%
This solution satisfies the initial condition $\bar{A}\left( 0\right) =0$
only for $c_{2}=0.$ The remaining irrational-power term is subleading at
small $z$ both for $L=0$ and $L=1$. For $L=2$, the general solution (induced
by the massless string mode) is%
\begin{equation}
\bar{A}_{S,L=2}\left( z\right) =-\frac{2}{3}\lambda ^{2}z^{2}-\frac{1}{24}%
\left( 1+c_{1}\right) \lambda ^{4}z^{4}+\frac{1}{6}\lambda ^{4}z^{4}\ln
\lambda z+c_{2}  \label{linL2}
\end{equation}%
where the initial condition $\bar{A}\left( 0\right) =0$ requires $c_{2}=0.$
For $L>2$, finally, the particular solution of the inhomogeneous equation is
identical to that for $L=0,1$ while the general solution of the homogeneous
equation differs. Their sum, 
\begin{align}
\bar{A}_{S,L>2}\left( z\right) & =\frac{L\lambda ^{2}z^{2}}{L^{2}-7}+\frac{%
\lambda ^{4}z^{4}}{2L^{2}-8}  \notag \\
& +c_{1}\lambda ^{2}z^{2}\cos \left( 2\sqrt{\frac{L^{2}-7}{3}}\ln \lambda
z\right) +c_{2}\lambda ^{2}z^{2}\sin \left( 2\sqrt{\frac{L^{2}-7}{3}}\ln
\lambda z\right) ,  \label{lL3}
\end{align}%
is the general solution of equation (\ref{A0lin}) and satisfies the initial
condition $\bar{A}\left( 0\right) =0$ for all (finite) values of $c_{1,2}$.
Note that for $L=0,1$ the coefficients of both $\lambda ^{2}z^{2}$ and $%
\lambda ^{4}z^{4}$ are negative, for $L=2$ the coefficient of $\lambda
^{2}z^{2}$ is negative and that of $\lambda ^{4}z^{4}$ can be chosen
positive (so that $\bar{A}_{L=0}>0$ for larger $z$), and for $L>2$ both
coefficients are positive.

The analysis of the linearized version%
\begin{equation}
-z^{2}A^{\prime \prime }-3zA^{\prime }+\frac{4}{3}\left( L^{2}-1\right) A-%
\frac{2}{3}\lambda ^{2}z^{2}\left( \lambda ^{2}z^{2}+2L\right) =0
\label{A1lin}
\end{equation}%
of equation (\ref{vde}) in the vector meson channel proceeds analogously.
For $L=0$ it yields the general solution%
\begin{equation}
\bar{A}_{V,L=0}\left( z\right) =-\frac{2}{7}\lambda ^{2}z^{2}-\frac{1}{38}%
\lambda ^{4}z^{4}+c_{1}\left( \lambda z\right) ^{-1}\cos \left( \frac{\ln
\lambda z}{\sqrt{3}}\right) +c_{2}\left( \lambda z\right) ^{-1}\sin \left( 
\frac{\ln \lambda z}{\sqrt{3}}\right)
\end{equation}%
where $\bar{A}_{V,L=0}\left( 0\right) =0$ demands $c_{1}=c_{2}=0$. For $L=1$
one finds%
\begin{equation}
\bar{A}_{V,L=1}\left( z\right) =-\frac{1}{3}\lambda^{2}z^{2}-\frac{1}{36}%
\lambda ^{4}z^{4}+c_{1}\frac{1}{\lambda^{2}z^{2}}+c_{2}
\end{equation}%
where $\bar{A}_{V,L=1}\left( 0\right) =0$ again requires $c_{1}=c_{2}=0$.
The solution for $L>1$, finally, is%
\begin{equation}
\bar{A}_{V,L>1}\left( z\right) =\frac{L\lambda ^{2}z^{2}}{L^{2}-7}+\frac{%
\lambda ^{4}z^{4}}{2L^{2}-38}+c_{1}\left( \lambda z\right) ^{-1-\sqrt{\frac{%
4L^{2}-1}{3}}}+c_{2}\left( \lambda z\right) ^{-1+\sqrt{\frac{4L^{2}-1}{3}}}
\label{lvl2}
\end{equation}%
where $\bar{A}_{V,L>1}\left( 0\right) =0$ demands $c_{1}=0$ while $c_{2}$
remains unconstrained. For $L=2$ the irrational power term provides the
leading small-$z$ behavior. The coefficient of the $\lambda ^{2}z^{2}$ ($%
\lambda ^{4}z^{4}$) term turns positive for $L\geq 3$ ($L\geq 4$).

The inhomogeneities in Eqs. (\ref{A0lin}) and (\ref{A1lin}) determine the
small-$z$ behavior of the solutions (with the exception of $\bar{A}_{V,L=2}$%
). In our context, this has two pertinent consequences. First, the leading
small-$z$ dependence is generally determined by the special solutions of the
inhomogeneous equation and therefore completely fixed, i.e. only the
subleading small-$z$ behavior depends on the boundary conditions.

A second useful consequence of the inhomogeneities in Eqs. (\ref{A0lin}) and
(\ref{A1lin}) - which are the same as those in Eqs. (\ref{mde}) and (\ref%
{vde}) - is related to the fact that the leading small-$z$ behavior of the
solutions to the full equations is identical to that of their linearized
counterparts. Hence at small $z$ the modulus of all solutions with $A\left(
0\right) =0$ grows as $\lambda ^{2}z^{2}$ (except for $\bar{A}_{S,L=0}$ and $%
\bar{A}_{V,L=2}$), partially with trigonometric or logarithmic corrections.

\end{document}